\documentclass{llncs}

\usepackage{enumerate}
\usepackage{upgreek}
\usepackage[english]{babel}
\usepackage{latexsym,amssymb,amsmath}
\usepackage{graphicx}
\usepackage{subfigure}
\usepackage{listings}
\usepackage{hyperref} 

\def\imagetop#1{\vtop{\null\hbox{#1}}}

\usepackage[hang,small,bf]{caption}
\newcommand{\W}{\mathcal{W}}

\newcommand{\R}{\mathbb{R}}
\renewcommand{\O}{\mathcal{O}}

\begin{document}
\title{Minimum Average Distance Triangulations}

\author{
L\'{a}szl\'{o} Kozma
}
\institute{Universit\"{a}t des Saarlandes, Saarbr\"{u}cken, Germany\\
\email{kozma@cs.uni-saarland.de}}

\maketitle

\begin{abstract}
We study the problem of finding a triangulation $T$ of a planar point set $S$ such as to minimize the expected distance between two points $x$ and $y$ chosen uniformly at random from $S$. By distance we mean the length of the shortest path between $x$ and $y$ along edges of $T$, with edge weights given as part of the problem. In a different variant of the problem, the points are vertices of a simple polygon and we look for a triangulation of the interior of the polygon that is optimal in the same sense.
We prove that a general formulation of the problem in which the weights are arbitrary positive numbers is strongly NP-complete. 
For the case when all weights are equal we give polynomial-time algorithms. In the end we mention several open problems.
\end{abstract}
\section{Introduction}

The problem addressed in this paper 
is a variant of the classical \emph{network design} problem. In many applications, the average routing cost between pairs of nodes is a sensible network characteristic, one that we seek to minimize. If costs are \emph{additive} (e.g., time delay) and \emph{symmetric}, an edge-weighted, undirected graph $G$ is a suitable model of the connections between endpoints. The task is then to find a spanning subgraph $T$ of $G$ that minimizes the average distance. Johnson \emph{et al.}~\cite{Johnson} study the problem when the total edge weight of $T$ is required to be less than a given budget constraint. They prove this problem to be NP-complete, even in the special case when all weights are equal and the budget constraint forces the solution to be a \emph{spanning tree}. 

Here we study the problem in a planar embedding: vertices of $G$ are points in the plane, edges of $G$ are straight segments between the points and weights are given as part of the problem. Instead of limiting the total edge weight of the solution, we require the edges of $T$ to be non-intersecting. 
From a theoretical point of view this turns out to be an essential difference: the problem now has a geometric structure that we can make use of. As an application we could imagine that we wanted to connect $n$ cities with an optimal railroad network using straight line connections and no intersections. We now give a more precise definition of the problem.

Given a set of points $S = \{p_1, \dots, p_n\} \subset \R^2$, and weights $w:S^2 \rightarrow \R$, having $w(x,x)=0$ and $w(x,y)=w(y,x)$, for all $x,y \in S$, we want to find a \emph{geometric}, \emph{crossing-free} graph $T$ with vertex set $S$ and edge weights given by $w$, such that the expected distance between two points chosen uniformly at random from $S$ is as small as possible. By \emph{distance} we mean the length of the shortest path in $T$ and we denote it by $d_T$. Since adding an edge cannot increase a distance, it suffices to consider \emph{maximal} crossing-free graphs, i.e., \emph{triangulations}. We call this the {\sc Minimum Average Distance Triangulation (madt)} problem.

The previous formulation, if we omit the normalizing factor, amounts to finding a triangulation $T$ that minimizes the following quantity:
\[
\mathcal{W}(T) = \displaystyle\sum_{1 \leq i < j \leq n}{d_T(p_i, p_j)}.
\]


Similarly, we ask for the triangulation $T$ of the interior of a polygon with $n$ vertices that has the minimum value $\mathcal{W}(T)$. In this case the triangulation consists of all boundary edges and a subset of the diagonals of the polygon.


We note that in mathematical chemistry, the quantity $\mathcal{W}(T)$ is a widely used characteristic of molecular structures, known as Wiener index \cite{Wiener1947,rouvray_}. 
Efficient computation of the Wiener index for special graphs, as well as its combinatorial properties have been the subject of significant research \cite{springerlink:10.1007/BF01167206,springerlink:10.1023/A:1010767517079,Nilsen08wienerindex}. 

\paragraph{Optimal triangulations.} Finding optimal triangulations with respect to various criteria has been intensively researched in the past decades \cite{aurenhammer,Bern92meshgeneration}. 
One particularly well-studied problem is \emph{minimum weight triangulation} ({\sc mwt}). For polygons, the solution of {\sc mwt} is found by the $\mathcal{O}(n^3)$ algorithm due to Gilbert \cite{gilbert} and Klincsek \cite{klincsek}, a classical example of dynamic programming. For point sets and Euclidean weights {\sc mwt} was proven to be NP-hard by Mulzer and Rote \cite{mwt}. For unit weights {\sc mwt} is trivial since all triangulations have the same cost. 

In contrast to both {\sc mwt} and budgeted network design \cite{Johnson}, {\sc madt} is interesting even for unit weights. In case of simple polygons, the problem is neither trivial, nor NP-hard. The algorithm we give in \textsection\,\ref{sec22} uses dynamic programming but it is much more involved than the $\O(n^3)$ algorithm for {\sc mwt}. Surprisingly, the ideas of the {\sc mwt} algorithm do not seem to directly carry over to the {\sc madt} problem, which, in fact, remains open for Euclidean weights, even for polygons. What makes our criterion of optimality somewhat atypical is that it is highly nonlocal. It is nontrivial to decompose the problem into smaller parts and known techniques do not seem to help.

\paragraph{Our results.}  
We study triangulations of point sets and of polygons. 
In the case of equal weights on all allowed edges, we assume w.l.o.g.\ that the weights are equal to one and we refer to the distance as \textit{link distance}. Using link distance, the solution is easily obtained when one point or one vertex can be connected to all the others. This is shown in \textsection\,\ref{sec21}. For the more general case of simple polygons (when no vertex can be connected to all other vertices) in \textsection\,\ref{sec22} we give an algorithm with a running time of $\O(n^{11})$ that uses dynamic programming. Our approach exploits the geometric structure of the problem, making a decomposition possible in this case.

For general point sets and arbitrary positive, symmetric weights (not necessarily obeying the triangle inequality), in \textsection\,\ref{sec23} we prove the problem to be strongly NP-complete, ruling out the existence of an efficient exact algorithm or of a fully polynomial time approximation scheme (FPTAS), unless P=NP. The hardness proof is a gadget-based reduction from {\sc Planar3SAT}. Again, the nonlocality of the cost function makes the reduction somewhat difficult, requiring a careful balancing between the magnitudes of edge weights and the size of the construction.

We leave the problem open in the case of Euclidean weights but we present the results of computer experiments for certain special cases in \textsection\,\ref{sec3}.

\section {Results}
\subsection{Link distance with one-point-visibility}  \label{sec21}

We call a point set $S$ \textit{one-point-visible} if one of the points $p \in S$ can be connected to all the points $q \in S$, where $p \neq q$, using straight segments that do not contain a point of $S$, except at endpoints. This condition is less restrictive than the usual \textit{generality} (no three points collinear). 
Similarly, we call a polygon \textit{one-vertex-visible} if one of the vertices can be connected to all others with diagonals or boundary edges of the polygon. The set of \textit{one-vertex-visible} polygons includes all convex polygons. 
A \emph{fan} is a triangulation in which one point or vertex (called the \emph{fan handle}) is connected to all other points or vertices.

\begin{theorem} 
\label{th1}
 For a one-vertex-visible polygon every fan triangulation has the same average distance and this is the smallest possible. For a one-point-visible point set every fan triangulation has the same average distance and this is the smallest possible.
\end{theorem}
\begin{proof}
The smallest possible distance of 1 is achieved for exactly $2n-3$ pairs of vertices in polygons and $3n-h-3$ pairs of points in point sets (with $h$ points on the convex hull), for every triangulation (these are the pairs that are connected with an edge and all triangulations of the same polygon or point set have the same number of edges). In a fan triangulation, all remaining pairs are at distance 2 from each other: the path between two vertices not connected with an edge can go via the fan handle. 
~\hfill\qed
\end{proof}



\subsection{Link distance in simple polygons} \label{sec22}
We now look at polygons that do not admit a fan triangulation. It would be desirable to decompose the problem and deal with the parts separately. The difficulty lies in the fact that when we are triangulating a smaller piece of the polygon, the decisions affect not just the distances within that piece but also the distances between external vertices. We need to do some bookkeeping of these global distances, but first we make some geometric observations.

Assume that an optimum triangulation $T$ has been found. We use a clockwise ordering of the vertices $p_1$ to $p_n$ and we denote by 
$p_d$ be the third vertex of the triangle that includes $p_1 p_n$. Let us visit the vertices from $p_1$ to $p_d$ in clockwise order (if $p_1 p_d$ is a boundary edge, then we have no other vertices in between). Let $p_a$ be the last vertex in this order such that $d_T(p_a, p_1) < d_T(p_a,p_d)$. There has to be such a vertex, since $p_1$ itself has this property and $p_d$ does not. Let $p_c$ be the first vertex for which $d_T(p_c, p_d) < d_T(p_c,p_1)$. Again, such a vertex clearly exists (in a degenerate case we can have $p_1=p_a$ or $p_c=p_d$ or both). Let $p_b$ denote the vertex (other than $p_n$) that is connected to both $p_1$ and $p_d$ (unless $p_1 p_d$ is on the boundary).

\vspace{-0.05in}
\begin{figure}[h!]
  \begin{center}
    \subfigure[]{\label{fig2a}\includegraphics[scale=0.22]{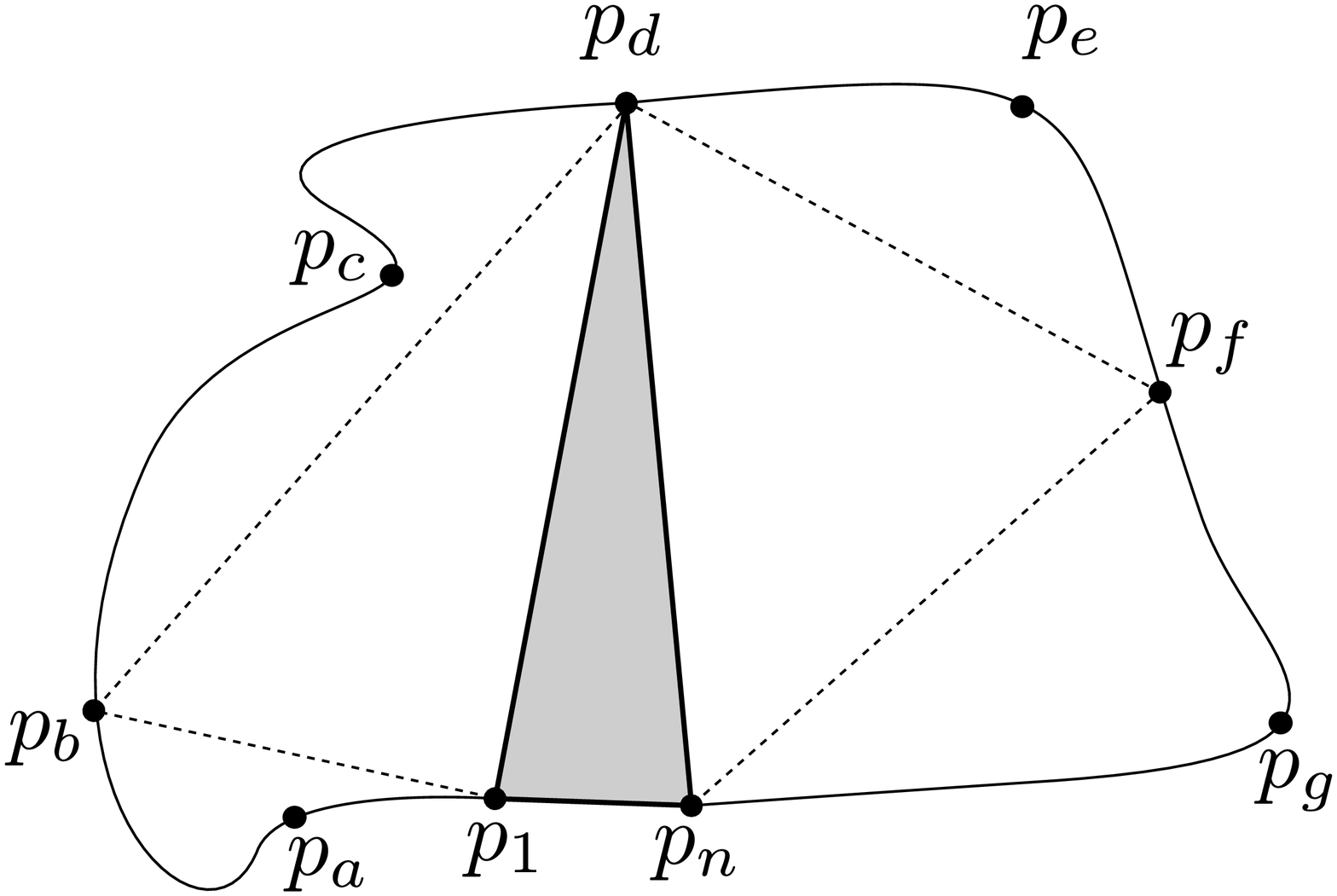}}
    \subfigure[]{\label{fig2b}\includegraphics[trim = -10mm 0mm 0mm 0mm, clip, scale=0.21]{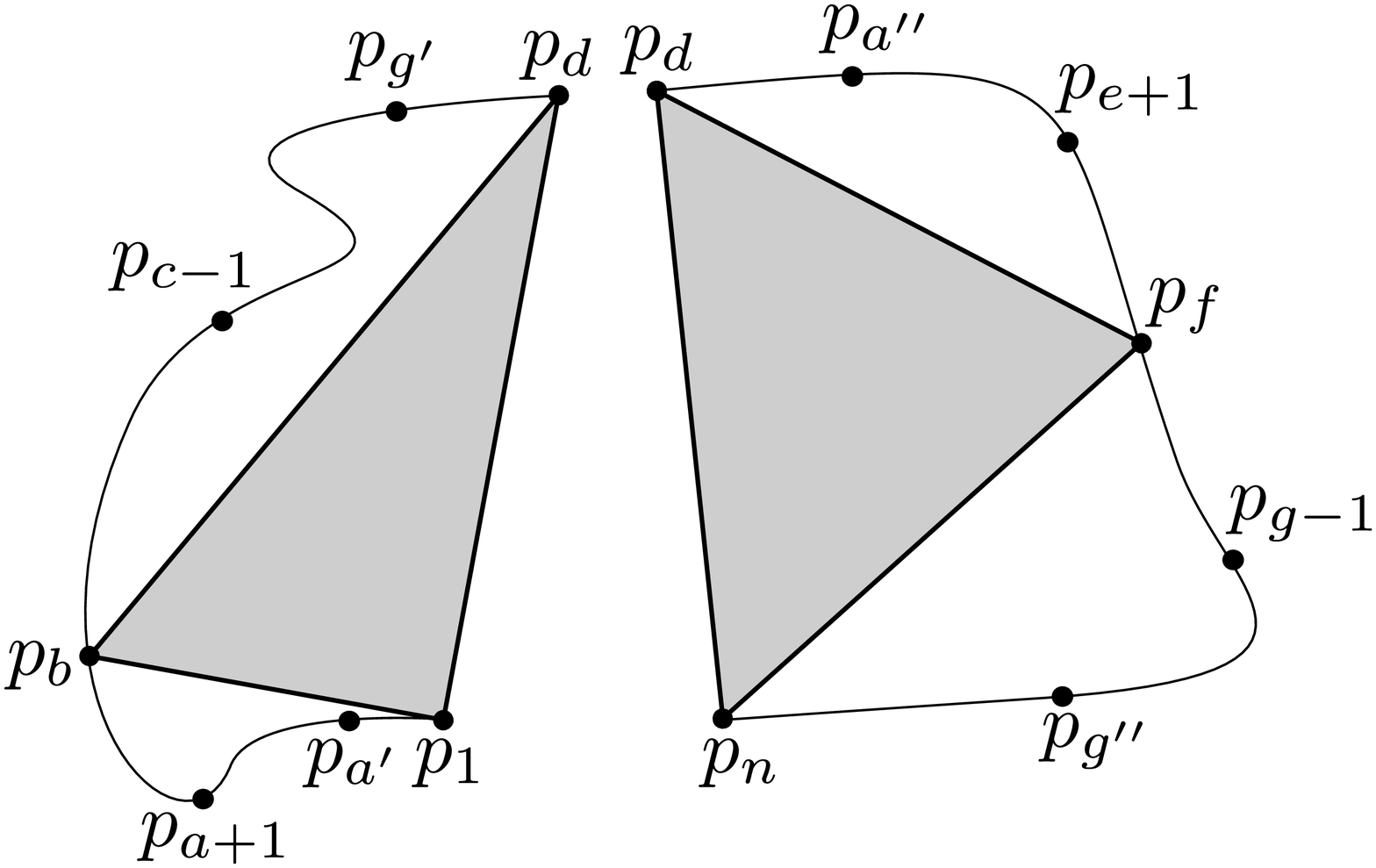}}
  \end{center}
  \caption{(a) Special vertices of the polygon. (b) Splitting up the polygon.}
  \label{fig2}
\end{figure}
\vspace{-0.15in}

On the other side of the triangle $p_1 p_d p_n$ we similarly visit the vertices from $p_d$ to $p_n$ and assign the label $p_e$ to the last vertex such that $d_T(p_e, p_d) < d_T(p_e,p_n)$ and $p_g$ to the first vertex such that $d_T(p_g, p_n) < d_T(p_g,p_d)$ and we let $p_f$ be the vertex connected to both $p_d$ and $p_n$ (Fig.~\ref{fig2a}). Now we can observe some properties of these vertices.

\begin{lemma} 
\label{lem1}
Let $1 \leq k \leq d$. Then the following hold (analogous statements hold for $d \leq k \leq n$):
\vspace{-0.05in}
\begin{enumerate}[(a)]
\item $d_T(p_k, p_1) < d_T(p_k,p_d)$ \, iff \, $1 \leq k \leq a$.
\item $d_T(p_k, p_1) > d_T(p_k,p_d)$ \, iff \, $c \leq k \leq d$.
\item $d_T(p_k, p_1) = d_T(p_k,p_d)$  iff \, $a < k < c$. In particular, if $p_b$ exists, then $a<b<c$. Otherwise $a=1$, $c=d=2$, and $p_1 p_2$ is on the boundary. 

\end{enumerate}

\vspace{-0.05in}
\end{lemma}
\begin{proof}
\emph{(a)} The largest index $k$ for which $d_T(p_k, p_1) < d_T(p_k, p_d)$ is $k=a$ by the definition of $p_a$. For the converse, observe that for all intermediary vertices $p_k$ on a shortest path between $p_1$ and $p_a$ we have $d_T(p_k, p_1) < d_T(p_k,p_d)$. 
Now suppose there is a vertex $p_l$, with $1 \leq l \leq a$, such that $d_T(p_l, p_d) \leq d_T(p_l,p_1)$. Such an inequality also holds for all intermediary vertices on the shortest path between $p_l$ and $p_d$. Since the shortest path between $p_l$ and $p_d$ intersects the shortest path between $p_a$ and $p_1$, the common vertex has to be at the same time strictly closer to $p_1$ and closer or equal to $p_d$, a contradiction.

\vspace{0.05in}
{\noindent}\emph{(b)} Similar argument as for \emph{(a)}.

\vspace{0.05in}
{\noindent}\emph{(c)} First, observe that $a<c$, otherwise some vertex would have to be strictly closer to both $p_1$ and $p_d$, a contradiction. Then, since for $1 \leq k < c$ we have $d_T(p_k, p_1) \leq d_T(p_k,p_d)$ and for $a<k\leq d$ we have $d_T(p_k, p_d) \leq d_T(p_k,p_1)$, it follows that for indices in the intersection of the two intervals ($a<k<c$) we have $d_T(p_k, p_d) = d_T(p_k,p_1)$. The converse follows from (a) and (b). Also, we have $d_T(p_b,p_1) = d_T(p_b,p_d) = 1$. ~\hfill\qed
\end{proof}

Equipped with these facts, we can split the distance between two vertices on different sides of the $p_1 p_d p_n$ triangle into locally computable components.

Let $1 \leq x \leq d$. Consider the shortest path between $p_x$ and $p_d$. Clearly, for all vertices $p_k$ on this path $1 \leq k \leq d$ holds, otherwise the path would go via the edge $p_1 p_n$ and it could be shortened via $p_1 p_d$. Similarly, given $d \leq y \leq n$, for all vertices $p_k$ on the shortest path between $p_y$ and $p_n$, we have $d \leq k \leq n$. We conclude that $d_T(p_x, p_d)$ and $d_T(p_y, p_n)$ only depend on the triangulations of $(p_1, \dots, p_d)\;$ and $\;(p_d, \dots, p_n)$ respectively. We now express the global distance $d_T(p_x, p_y)$ in terms of these two local distances.

\begin{lemma} 
\label{lem2}
Let $p_1,\dots,p_n$ defined as before, $1 \leq x \leq d$, and $d \leq y \leq n$, and let $\upphi = d_T(p_x, p_d) + d_T(p_y, p_n)$. Then the following holds, covering all possible values of $x$ and $y$:
\[ d_T(p_x, p_y) = \left\{ \begin{array}{lll}
         \upphi-1 & \mbox{ if $d \leq y \leq e$};\\
         \upphi+1 & \mbox{ if $g \leq y \leq n\;$ and $\;a < x \leq d$};\\
         \upphi & \mbox{ otherwise }.\end{array} \right. \]
\end{lemma}
\begin{proof}
In each of the cases we use Lemma~\ref{lem1} to argue about the possible ways in which the shortest path can cross the triangle $p_1 p_d p_n$.
For example, if $d \leq y \leq e$, the shortest path goes through $p_d$, therefore we have $d_T(p_x,p_y) = d_T(p_x,p_d)+d_T(p_y,p_d)$. Since $d_T(p_y,p_d) = d_T(p_y,p_n) - 1$, we obtain $d_T(p_x, p_y) = \upphi-1$. The other cases use similar reasoning and we omit them for brevity. ~\hfill\qed

\end{proof}

Lemma~\ref{lem2} allows us to decompose the problem into parts that can be solved separately. We proceed as follows: we \textit{guess} a triangle $p_1 p_d p_n$ that is part of the optimal triangulation and we use it to split the polygon in two. We also \emph{guess} the special vertices $p_a, p_c, p_e, p_g$. We recursively find the optimal triangulation of the two smaller polygons with vertices $(p_1,\dots,p_d)$ and $(p_d,\dots,p_n)$. Besides the distances \textit{within} the subpolygons we also need to consider the distances \textit{between} the two parts. Using Lemma~\ref{lem2} we can decompose these distances into a distance to $p_d$ in the left part, a distance to $p_n$ in the right part and a constant term. 

We now formulate an extended cost function $\mathcal{W}_\textrm{EXT}$, that has a second term for accumulating the distances to endpoints that result from splitting up global distances. The coefficient $\upalpha \in \mathbb{N}$ will be uniquely determined by the sizes of the polygons, which in turn are determined by the choice of the index $d$. We express this new cost function for a general subpolygon $(p_i, \dots, p_j)$:
\[
\mathcal{W}_\textrm{EXT}(T, \upalpha) \Big |_{i}^j = \displaystyle\sum_{i \leq x < y \leq j}{d_T(p_x, p_y)} + \upalpha \sum_{i \leq x \leq j}{d_T(p_x, p_j)}.
\]
Observe that minimizing $\mathcal{W}_\textrm{EXT}(T, 0) \Big |_{1}^n$ solves the initial problem. Using Lemma~\ref{lem2} we can split the sums and the distances until we can express $\mathcal{W}_\textrm{EXT}$ recursively in terms of smaller polygons and the indices of special vertices $a, c, e, g$. Note that $p_a$, $p_c$, $p_e$, $p_g$ play the same role as in the earlier discussion, but now the endpoints are $p_i$ and $p_j$ instead of $p_1$ and $p_n$.

\vspace{-0.15in}
\begin{align*}
&\mathcal{W}_\textrm{EXT}(T, \upalpha) \Big |_{i}^j  = 
 \displaystyle\sum_{i \leq x < y \leq d}{d_T(p_x,p_y)} + \sum_{d \leq x < y \leq j}{d_T(p_x,p_y)}+ \sum_{\substack{i \leq x \leq d\\d \leq y \leq j}}{d_T(p_x,p_y)}\\
& 
 - \sum_{i \leq x \leq j}{d_T(p_x,p_d)} + \upalpha \sum_{i \leq x \leq d}{d_T(p_x, p_j)} + \upalpha \sum_{d \leq x \leq j}{d_T(p_x, p_j)} - \upalpha \cdot d_T(p_d, p_j)\\
& \quad \quad \quad \quad \, \, \, \, \, \, \quad = \ \mathcal{W}_\textrm{EXT}(T, \upalpha + j-d) \Big |_{i}^d + \mathcal{W}_\textrm{EXT}(T, \upalpha + d-i) \Big |_{d}^j\\
& \quad \quad \quad \quad \quad \quad \quad + (\upalpha + j-g+1)(d-a-1) + (e-d+1)(i-d).\\
\end{align*}
\vspace{-0.35in}


How can we make sure that the constraints imposed by the choice of the special vertices $p_a$, $p_c$, $p_e$, $p_g$ are respected by the recursive subcalls? If the left side of the triangle is on the boundary ($d=i+1$), it follows that $a=i$ and $c=d$ and it is trivially true that $d_T(p_a,p_i) < d_T(p_a,p_d)$. Similarly, if $d=j-1$, it follows that $e=d$ and $g=j$, therefore $d_T(p_e,p_d) < d_T(p_e,p_j)$. The general case remains, when one of the sides of the triangle is not on the boundary. The following lemma establishes a necessary and sufficient condition for the constraints to hold. We write it only for the side $p_i p_d$ and special indices $a$ and $c$, a symmetric argument works for the side $p_d p_j$ and special indices $e$ and $g$.

\begin{lemma} 
\label{lem3}
Let $p_i,\dots,p_j$ be a triangulated polygon. Assume that the triangulation contains the triangles $p_i p_d p_j$ and $p_i p_b p_d$. Then the following hold:
\vspace{-0.05in}
\begin{enumerate}[(a)]
\item We have $a$ as the largest index ($i \leq a \leq d$) for which $d_T(p_a,p_i) < d_T(p_a,p_d)$ iff $a+1$ is the smallest index ($i \leq a+1 \leq b$) such that $d_T(p_{a+1},p_b) < d_T(p_{a+1},p_i)$.
\item We have $c$ as the smallest index ($i \leq c \leq d$) for which $d_T(p_c,p_d) < d_T(p_c,p_i)$ iff $c-1$ is the largest index ($b \leq c-1 \leq d$) such that $d_T(p_{c-1},p_b) < d_T(p_{c-1},p_d)$.
\end{enumerate}
\end{lemma}
\begin{proof}
\emph{(a)} If $a$ is the largest index such that $d_T(p_a,p_i) < d_T(p_a,p_d)$ then $a+1$ is the smallest index such that $d_T(p_{a+1},p_d) \leq d_T(p_{a+1},p_i)$. Since $a+1 \leq b$, the shortest path between $p_{a+1}$ and $p_d$ contains $p_b$, therefore $d_T(p_{a+1},p_b) < d_T(p_{a+1},p_i)$. To see that $a+1$ is the smallest index with this property, we need to prove that $d_T(p_k,p_b) \geq d_T(p_k,p_i)$ for all $i \leq k \leq a$. This inequality follows from $d_T(p_k,p_d)>d_T(p_k,p_i)$ and $d_T(p_k,p_d) = d_T(p_k, p_b)+1$. 

For the converse, assume $a+1$ to be the smallest index such that $d_T(p_{a+1},p_b) < d_T(p_{a+1},p_i)$. Then $d_T(p_a,p_b) \geq d_T(p_a,p_i)$. Since $d_T(p_a,p_d) = d_T(p_a,p_b) + 1$, it follows that $d_T(p_a,p_i) < d_T(p_a,p_d)$. To see that $a$ is the largest index with this property, we need $d_T(p_{k},p_i) \geq d_T(p_{k},p_d)$ for all $a<k \leq b$ (for $k > b$ the inequality clearly holds). Again, this follows from $d_T(p_{k},p_i) > d_T(p_{k},p_b)$ and $d_T(p_{k},p_d) = d_T(p_{k},p_b) + 1$. \\
\vspace{0.15in}
\emph{(b)} Similarly. ~\hfill\qed
\vspace{-0.05in}
\end{proof}

Procedure \textsc{ext} (see Fig.~\ref{figcode}) returns the cost $\W_\textrm{EXT}$ of a triangulation that minimizes this cost. 
We can modify our procedure without changing the asymptotic running time, such as to return the actual triangulation achieving minimum cost. The results are then merged to form the full solution. 

Lemma~\ref{lem3} tells us that in all recursive calls two of the four special vertices are fixed and we only need to guess the remaining two. We label these new vertices as $p_a'$, $p_g'$ and $p_a''$, $p_g''$. They play the same role in their respective small polygons as $a$ and $g$ in the large polygon (see Fig.~\ref{fig2b} for illustration). The notation $p \leftrightarrow q$ indicates the condition that vertices $p$ and $q$ see each other within the polygon.
\vspace{-0.1in}
\begin{figure}[h!]

\rule{\linewidth}{0.2mm}

\vspace{0.04in}
\small{

\textbf{procedure} \textsc{ext}\,$\bigl((p_i, \dots, p_j),\; p_a, p_c, p_e, p_g,\; \upalpha \bigr)$:

\ \ \ \ \ \ \ \textbf{if} $(a=i)$ and $(c=e=i+1)$ and $(g=j=i+2)$:

\ \ \ \ \ \ \ \ \ \ \ \textbf{return} $3 + 2\upalpha$;  \ \ \ \ \ \   /* \textit{the polygon has only three vertices} */

\ \ \ \ \ \ \ \textbf{else}:

\begin{tabular}{llp{15cm}}
\imagetop{$\;\;\;\;\;\;\;\;\;\;\;\textrm{\textbf{return}} \displaystyle\min_{\substack{p_d, p_a', p_g', p_a'', p_g'':\\i \leq a' \leq
 a+1 \leq c-1 \leq
 g' \leq d\\d \leq a'' \leq 
e+1 \leq g-1 \leq
g'' \leq j\\ p_i \leftrightarrow p_d \leftrightarrow p_j}}{}$} &\imagetop{\parbox{15cm}{ $ \Big\{ \textsc{ext}\,\bigl((p_i, \dots, p_d), \;p_a', p_{a+1}, p_{c-1}, p_g', \;\upalpha + j-d \bigr)$ \\ $\;+\;\textsc{ext}\,\bigl((p_d, \dots, p_j),\; p_a'', p_{e+1}, p_{g-1}, p_g'',\; \upalpha + d-i \bigr) \\ + \;(\upalpha + j-g+1)(d-a-1) + (e-d+1)(i-d) \Big\} $;}}
\end{tabular}

\vspace{0.04in}

}
\rule{\linewidth}{0.2mm}

\caption{Procedure for finding the triangulation that minimizes $\mathcal{W}_\textrm{EXT}(T, \upalpha)$.}\label{figcode}
\end{figure}

\vspace{-0.15in}
The process terminates, since every subcall is on polygons of smaller sizes and we exit the recursion on triangles. Clearly every triangulation can be found by the algorithm and the correctness of the decomposition is assured by Lemma~\ref{lem2}. The fact that indices $a, c, e, g$ indeed fulfill their necessary role (and thus the expressions in the cost-decomposition are correct) is guaranteed by Lemma~\ref{lem3}. 

For the cases when there is no suitable vertex that can be connected to the required endpoints and whose index fulfills the required inequalities, we adopt the convention that the minimum of the empty set is $+\infty$, 
thus abandoning those branches in the search tree.

\begin{theorem} 
\label{thm2}
The running time of {\sc ext} on a polygon with $n$ vertices is 
$\O(n^{11})$.
\end{theorem}
\begin{proof}
Observe that all polygons in the function calls have contiguous indices, therefore we can encode them with two integers between 1 and $n$. Furthermore, if the initial call has $\upalpha=0$, then it can be shown that on each recursive call the parameter $\upalpha$ becomes ``$n$ minus the number of vertices in the polygon''. For this reason it is superfluous to pass $\upalpha$ as an argument. There are four remaining parameters which can all take $n$ different values. We can build up a table containing the return values of all $\O(n^6)$ possible function calls, starting from the smallest and ending with the full polygon. In computing one entry we take the minimum of $\O(n^5)$ values, giving a total running time of $\O(n^{11})$. ~\hfill\qed
\end{proof}


\subsection{Arbitrary positive weights} \label{sec23}

We now prove that the decision version of {\sc madt} for point sets is NP-complete if the edge weigths $w$ are arbitrary positive numbers, such that $w(p_i,p_j) = 0$ iff $i=j$ and ${w(p_i,p_j)=w(p_j,p_i)}$ for all $i,j$. Such a weight function is called
a \emph{semimetric}. 
We leave open the status of the problem for \emph{metric} weights (that obey the triangle inequality). 
We defer most of the details of the proofs in this section to the Appendix. 
Where possible, we give a short, intuitive explanation.\\

\textbf{{\sc madt}} (decision version): \emph{For given $\mathcal{W}^\star \in \mathbb{R}$, is there a triangulation $T$ of a given point set $S$ and weights $w$, such that $\mathcal{W}(T) \leq \mathcal{W}^\star$} ?\\

The problem is clearly in NP, since for a given triangulation $T$, we can use an all-pairs shortest path algorithm to compute $\mathcal{W}(T)$ and compare it with $\mathcal{W}^\star$ in polynomial time.

We prove NP-hardness using a reduction from {\sc Planar3SAT} \cite{DBLP:journals/siamcomp/Lichtenstein82}. In {\sc 3SAT}, given a 3-CNF formula, we ask whether there exists an assignment of truth values to the variables such that each clause has at least one \emph{true} literal. {\sc Planar3SAT} restricts the question to planar formulae: those that can be represented as a planar graph in which vertices represent both variables and clauses of the formula and there is an edge between clause $C$ and variable $x$ iff $C$ contains either $x$ or $\lnot x$.

Knuth and Ragunathan \cite{Knuth:1992:PCR:141099.141159} observed that {\sc Planar3SAT} remains\linebreak NP-complete if it is restricted to formulae embedded in the following fashion: variables are arranged on a horizontal line with three-legged clauses on the two sides of the line. Clauses and their three legs are properly nested, i.e., none of the legs cross each other. We can clearly have an actual embedding in which all three legs of a clause are straight lines and the ``middle'' leg is perpendicular to the line of the variables (Appendix: Fig.~\ref{fig6}). For simplicity, let us call such an embedding of a formula a \emph{planar circuit}.

We put two extra conditions on the admissible planar circuits such that {\sc Planar3SAT} remains NP-complete when restricted to planar circuits obeying these conditions: (\textbf{R1}) no variable appears more than once in the same clause, and (\textbf{R2}) every variable appears in at least two clauses.

\begin{lemma} 
\label{lema1}
Given a planar circuit $\upphi_1$, we can transform it into a planar circuit $\upphi_2$ that obeys \emph{R1} and \emph{R2}, such that $\upphi_2$ has a satisfying assignment iff $\upphi_1$ does.
\end{lemma}
\begin{proof}
We examine every possible way in which R1 or R2 can be violated and give transformations that remove the violations while preserving the planar embedding, as well as the satisfiability of the circuit.
~\hfill\qed
\end{proof}

\begin{figure}[h!]
  \centering
\includegraphics[scale=0.32]{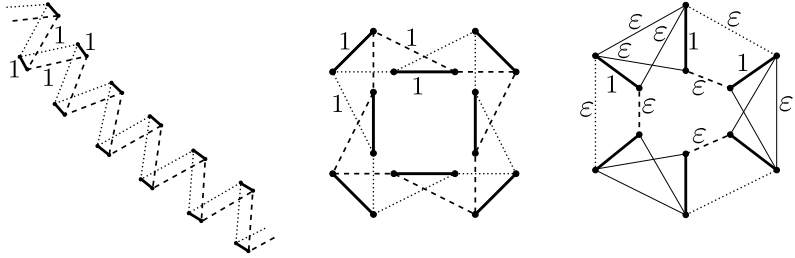}
  \caption{(a) Wire gadget. \quad (b) Simplified variable gadget. \quad (c) Clause gadget.} \label{fig4}
\end{figure}

\begin{figure}[h!]
  \centering
\includegraphics[scale=0.17]{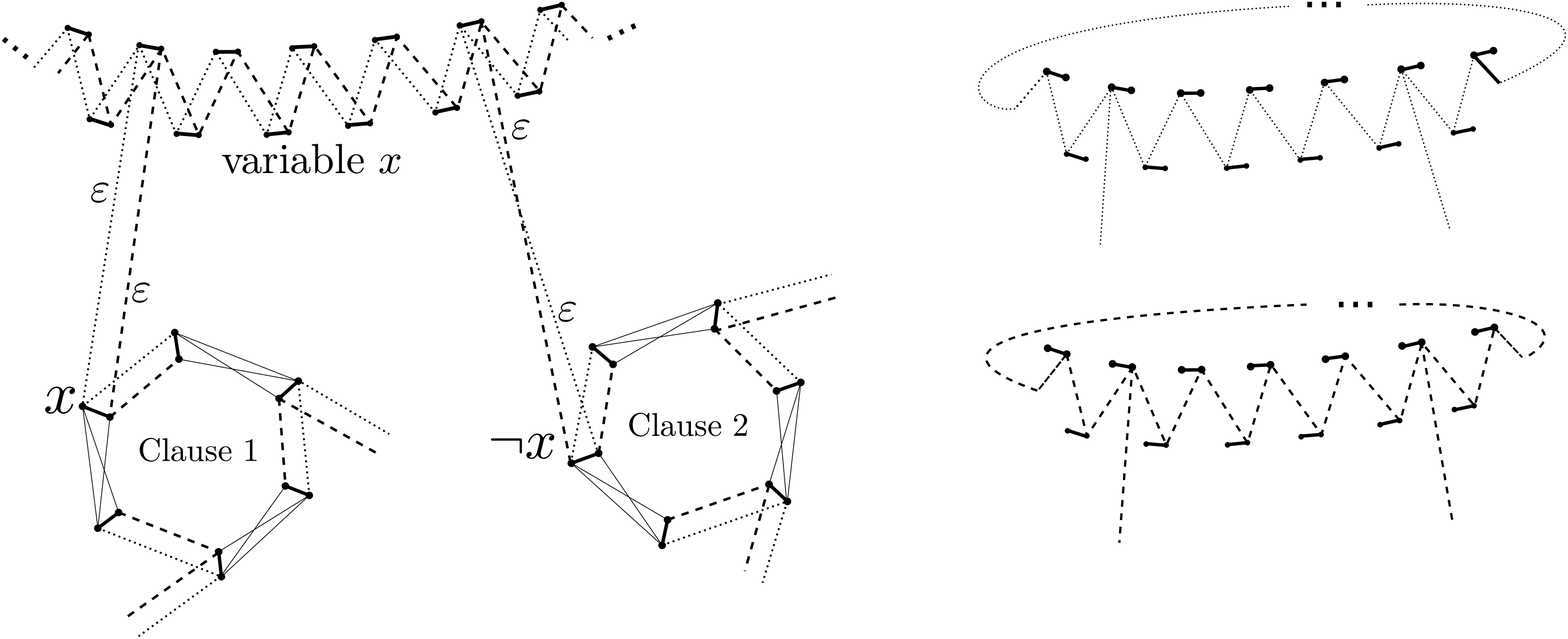}
  \caption{(a) Bridge between variable and clause. \quad (b) Pure triangulations of a variable.} \label{fig5}
\end{figure}

The gadgets used in the reduction are shown in Fig.~\ref{fig4}. They consist of points in the plane and the weights of the potential edges between them. Weights can take one of three values: the value 1, a small value $\varepsilon$ and a large value $\sigma$ (higher than any distance in the triangulation). We call edges with weight $\sigma$ \emph{irrelevant} and we do not show them in the figures. Including irrelevant edges in a triangulation never decreases the cost $\mathcal{W}$, therefore we can safely ignore them. The values $\varepsilon$ and $\sigma$ depend on the problem size (number of clauses and variables).

The basic building block is the \emph{wire}, shown in Fig.~\ref{fig4}(a). The thick solid edges are part of all triangulations.
From each pair of intersecting dotted and dashed edges exactly one is included in any triangulation. 
The weights of all non-irrelevant edges in a wire are 1. The wire-piece can be bent and stretched freely as long as we do not introduce new crossings or remove a crossing between two edges (irrelevant edges do not matter).

The \emph{variable} is a wire bent into a loop, with the corresponding ends glued together. We illustrate this in Fig.~\ref{fig4}(b) using a wire-piece with 16 vertices. The construction works with any $4k$ vertices for $k \geq 3$ and in fact we will use \emph{much} more than 16 vertices in each variable. 

The \emph{clause} gadget and the weights of its edges are shown in Fig.~\ref{fig4}(c). 

A \emph{bridge} is a pair of edges that links a variable to a clause.  A clause gadget has three fixed ``places'' where bridges are connected to it. We use \textit{parallel} or \textit{crossing} bridges, as seen in Fig.~\ref{fig5}(a). Given a {\sc Planar3SAT} instance (a planar circuit), we transform it into an instance of {\sc madt} as follows: we replace the vertices of the planar circuit by variable- or clause gadgets and we replace the edge between clause $C$ and variable $x$ by a parallel bridge if $C$ contains $x$ and by a crossing bridge if $C$ contains $\lnot{x}$.





\begin{lemma} 
\label{lema1_}
Using the gadgets and transformations described above we can represent any planar circuit as a {\sc madt} instance. ~\hfill\qed
\end{lemma}

Since the gadgets allow a large amount of flexibility, the proof is straightforward. 
Now we can formulate our main theorem:

\begin{theorem} 
\label{thm3}
We can transform any planar circuit $\phi$ into a {\sc madt} instance consisting of a point set $S$ in the plane, a \emph{semimetric} weight function $w: S^2 \rightarrow \mathbb{R}$ and a threshold $\;\mathcal{W}^\star$, such that $S$ admits a triangulation $T$ with $\mathcal{W}(T) \leq \mathcal{W}^\star$ iff $\phi$ has a satisfying assignment. All computations can be done in polynomial time and the construction is of polynomial size, as are all parameters.
\end{theorem}

\begin{corollary}
{\sc madt} with semimetric weights is strongly NP-complete.
\end{corollary}

The proof of Theorem~\ref{thm3} relies on a sequence of lemmas. The high level idea is the following: we call a triangulation of the construction \emph{pure}, if every variable gadget, together with its associated bridges contains either only dashed edges or only dotted edges (besides the thick solid edges), see Fig.~\ref{fig5}(b). First we show that we only need to consider \emph{pure} triangulations and thus we can use the pure states of the gadgets to encode an assignment of truth values, with the convention \emph{dotted}\,$\rightarrow(true)$, and \emph{dashed}\,$\rightarrow(false)$. Then, we prove that satisfying assignments lead to triangulations with the smallest cost. Finally, we bound the difference in cost between different satisfying assignments and we show how to generate a \emph{baseline} triangulation, with cost not far from the cost of a satisfying assignment (if one exists). The cost of the baseline triangulation can be computed in polynomial time. 

We denote the number of variables in our planar circuit by $n_v$ and the number of clauses by $n_c$. Due to condition R2, we have $n_v \leq 1.5 n_c$. We denote the number of vertices in a variable gadget between two bridges (not including the bridge endpoints) by $N$ (the same value for all variables). In Fig.~\ref{fig5}(a), for instance, we have $N=14$. The proof requires a careful balancing of the parameter $N$ describing the size of the construction and the weight $\varepsilon$.




\begin{lemma} 
\label{thm5}
If $N > 5 \cdot 10^5 {n_c}^3$, for any impure triangulation $T_\textrm{impure}$ of the construction we can find a pure triangulation $T_\textrm{pure}$ such that $\mathcal{W}(T_\textrm{pure})<\mathcal{W}(T_\textrm{impure})$.
\end{lemma}

\begin{proof}

The main idea is that if a variable is impure then the loop of the gadget is necessarily broken in some place and this leads to a penalty in cost that cannot be offset by any other change in the triangulation. ~\hfill\qed
\end{proof}


\begin{lemma} 
\label{thm6}
Let $T_\textrm{SAT}$ be a triangulation corresponding to a satisfying assignment of a planar circuit (assuming such an assignment exists) and let $T_\textrm{nonSAT}$ be the triangulation with smallest cost $\mathcal{W}(T_\textrm{nonSAT})$ among all triangulations corresponding to nonsatisfying assignments. Then, for $N > 5 \cdot 10^5 {n_c}^3$ and $\varepsilon < \frac{1}{N^2}$, we have $\mathcal{W}(T_\textrm{nonSAT}) - \mathcal{W}(T_\textrm{SAT}) \geq \frac{N^2}{32}$. ~\hfill\qed
\end{lemma}


\begin{lemma} 
\label{thmxy}
If $T_\textrm{SAT1}$ and $T_\textrm{SAT2}$ are two triangulations corresponding to different satisfying assignments, then, given previous bounds on $N$ and $\varepsilon$, we have that $|\mathcal{W} (T_\textrm{SAT1}) - \mathcal{W} (T_\textrm{SAT2})| \leq 150 {n_c}^2 N$. ~\hfill\qed
\end{lemma}





\begin{lemma} 
\label{thmxyz}
For any $T_\textrm{baseline}$ and any $T_\textrm{SAT}$, given previous bounds, we have $|\mathcal{W}(T_\textrm{baseline})-\mathcal{W}(T_\textrm{SAT})| \leq 150 {n_c}^2 N$. ~\hfill\qed
\end{lemma}








We can now generate the threshold as $\mathcal{W}^\star = \mathcal{W}(T_\textrm{baseline}) + 300 {n_c}^2 N + 1$. In accordance with our previous constraints we set $N= 10^6 {n_c}^3$, $\varepsilon=\frac{1}{10^{13} {n_c}^6}$, and this ensures that $W^\star$ falls in the gap between satisfying and nonsatisfying triangulations.
We note that all constructions and computations can be performed in polynomial time and all parameters have polynomial magnitude. This concludes the proof of NP-completeness. ~\hfill\qed

\section{Open questions} \label{sec3}
The main unanswered question is the status of the problem for metric, in particular Euclidean weights. For polygons we have not succeeded in establishing results similar to Lemma~\ref{lem1} that would enable a dynamic programming approach. For point sets we suspect that the problem remains NP-hard, but we have not found a reduction.

\begin{figure}[h!]
  \centering
    \subfigure[]{\label{fig23a}\includegraphics[scale=0.24]{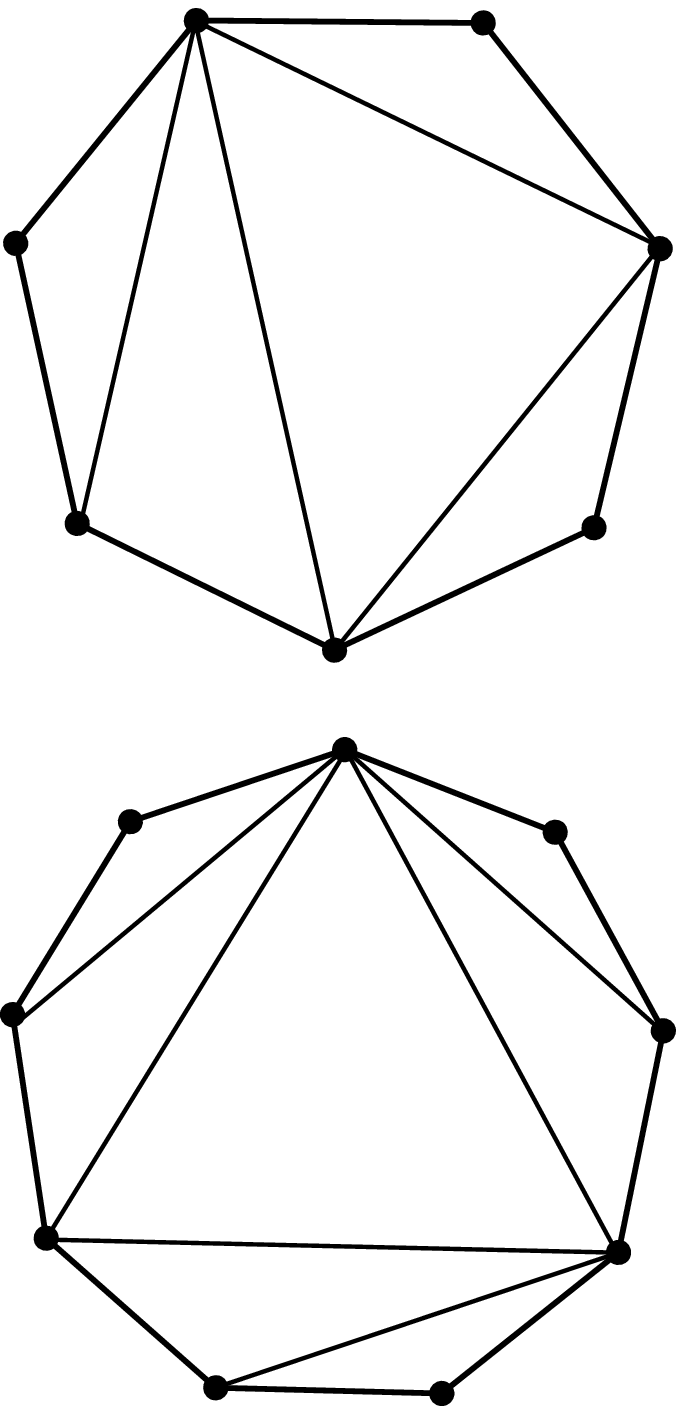}}
    \subfigure[]{\label{fig23b}\includegraphics[trim = -50mm 0mm 0mm 0mm, clip, scale=0.24]{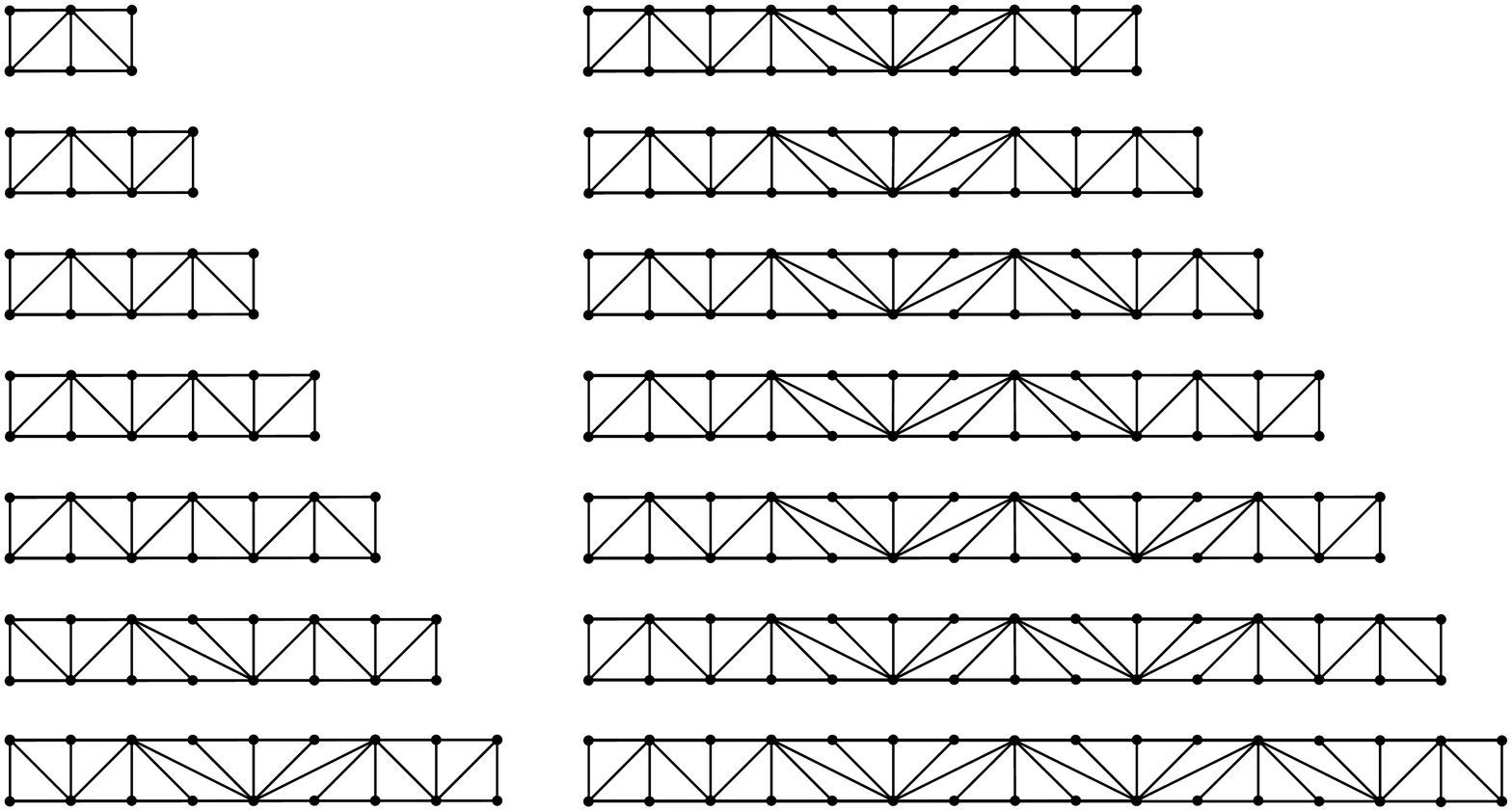}}
  \caption{(a) Solution for regular polygons. \quad (b) Solution for 2-by-n grids.} \label{figexa}
\end{figure}

The Euclidean-distance problem remains open even for the special case of regular polygons with $n$ vertices (in this case the boundary of the polygon already gives a $\frac{\pi}{2}$-approximation). Computer simulation shows that the exact solution is the fan for $n$ up to 18, \emph{except} for the cases $n=7$ and $n=9$, where the solutions are shown in Fig.~\ref{fig23a}. We conjecture these to be the only counterexamples. As another special case, Fig.~\ref{fig23b} shows the solutions obtained for a 2-by-n grid, for $n$ up to 16.

The problem remains open with unit weights in the case of point sets not admitting a fan, even in special cases such as a 3-by-n grid.
For the NP-hard variant of the problem the question remains whether a polynomial-time approximation scheme (PTAS) exists (an FPTAS is impossible, unless P=NP).

Variants of the problem that we have not studied include placing various constraints on the allowed geometric graphs \emph{besides} non-crossing, such as a total budget on the sum of edge weights or bounded vertex-degree, as well as the case when Steiner points are allowed. One can also study the problem of maximizing the average distance, instead of minimizing it.

\section{Acknowledgement}
I thank my advisor, Raimund Seidel, for mentioning the problem at the INRIA-McGill-Victoria Workshop on Computational Geometry (2011) at the Bellairs Research Institute, and for valuable comments. I also thank several anonymous reviewers for pointing out errors and suggesting improvements.

\bibliographystyle{splncs}
\bibliography{myrefs}	
\newpage

\section{Appendix}

\paragraph{Illustration of Theorem~1.} ~\\

\begin{figure}[h!]
  \begin{center}
    \subfigure[]{\label{fig:fig00a}\includegraphics[scale=0.5]{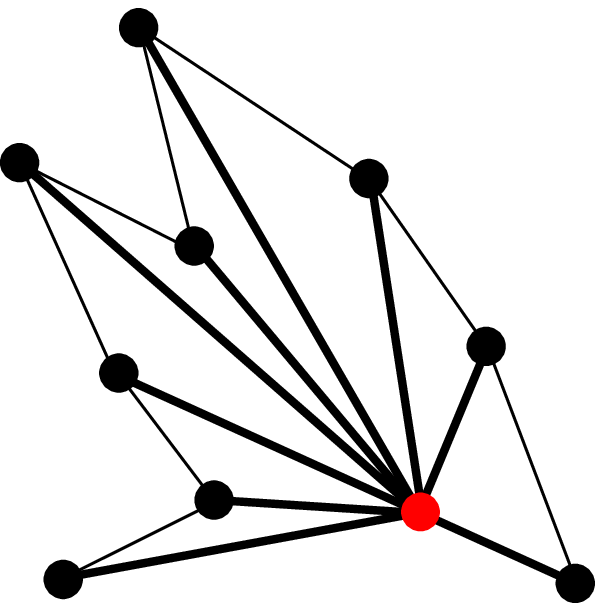}}
    \subfigure[]{\label{fig:fig00b}\includegraphics[scale=0.5]{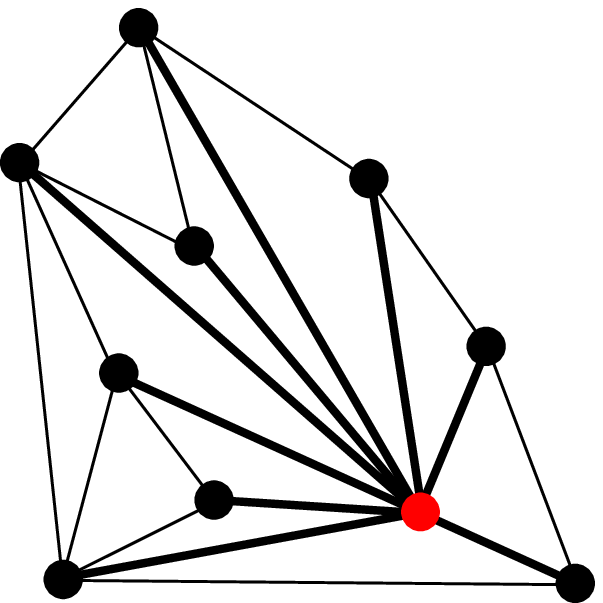}}
  \end{center}
  \caption{(a) Fan triangulation of a polygon. (b) Fan triangulation of a point set.}
  \label{fig00}
\end{figure}

\paragraph{Details of the splitting up of the cost-function using Lemma~\ref{lem2}:}
\begin{align*}
\mathcal{W}_\textrm{EXT}(T, \upalpha) \Big |_{i}^j & = \displaystyle\sum_{i \leq x < y \leq j}{d_T(p_x, p_y)} + \upalpha \sum_{i \leq x \leq j}{d_T(p_x, p_j)}\\
& = \displaystyle\sum_{i \leq x < y \leq d}{d_T(p_x,p_y)} + \sum_{d \leq x < y \leq j}{d_T(p_x,p_y)} - \sum_{d \leq x \leq j}{d_T(p_x,p_d)} + \sum_{\substack{i \leq x \leq d\\d \leq y \leq j}}{d_T(p_x,p_y)}\\
& \quad - \sum_{i \leq x \leq d}{d_T(p_x,p_d)} + \upalpha \sum_{i \leq x \leq d}{d_T(p_x, p_j)} + \upalpha \sum_{d \leq x \leq j}{d_T(p_x, p_j)} - \upalpha \cdot d_T(p_d, p_j)\\
& = \ \mathcal{W}_\textrm{EXT}(T, 0) \Big |_{i}^d + \mathcal{W}_\textrm{EXT}(T, 0) \Big |_{d}^j - \left( \sum_{d \leq x \leq j}{d_T(p_x,p_j)} - (e-d+1) + (j-g+1) \right)\\
& \quad + \left(\sum_{\substack{i \leq x \leq d\\d \leq y \leq j}}{ \bigl( d_T(p_x,p_d) + d_T(p_y,p_j) \bigr)} - (d-i+1)(e-d+1) + (j-g+1)(d-a) \right)\\
& \quad - \sum_{i \leq x \leq d}{d_T(p_x,p_d)} + \upalpha \left( \sum_{i \leq x \leq d}{d_T(p_x, p_d)} + (d-a) \right) + \upalpha \sum_{d \leq x \leq j}{d_T(p_x, p_j)} - \upalpha \\
& = \ \mathcal{W}_\textrm{EXT}(T, \upalpha + j-d) \Big |_{i}^d + \mathcal{W}_\textrm{EXT}(T, \upalpha + d-i) \Big |_{d}^j\\
& \quad + (\upalpha + j-g+1)(d-a-1) + (e-d+1)(i-d).\\
\end{align*}

\newpage

\begin{proof} (Lemma~\ref{lema1})

First we give the transformations that establish conditions R1 and R2 while preserving the satisfiability of the formula. We use the following notation: $x$ and $y$ are variables of the original formula, $a$, $b$, $c$ and $d$ are variables introduced during the transformation. A clause $(x \lor y \lor z)$ is denoted simply as $(xyz)$ and $\emptyset$ means no clause. The values \emph{True} and \emph{False} are denoted by \emph{T} and \emph{F}, respectively.

\vspace{-0.15in}
\begin{figure}[h!]
  \centering
\includegraphics[scale=0.2]{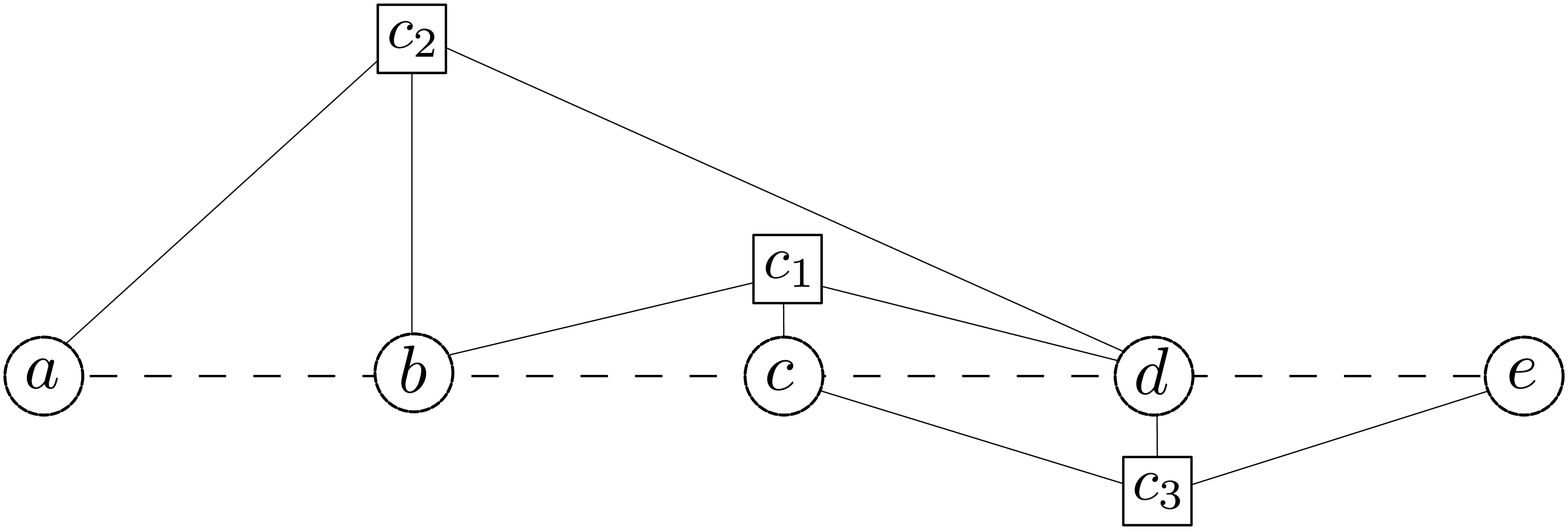}
  \caption{Planar circuit of the formula $(b \lor \lnot{c} \lor \lnot{d})\wedge(a \lor \lnot{b} \lor d)\wedge(\lnot{c} \lor d \lor e)$.} \label{fig6}
\end{figure}
\vspace{-0.15in}

The following replacements cover all possible scenarios in which a variable is repeated:
\begin{align*}
&(x x x) \rightarrow (xab) \land (x \lnot a b) \land (x \lnot b c) \land (x \lnot b \lnot c) \tag{$x=T$}\\
&(x x \lnot{x}) \rightarrow \emptyset \tag{$x$ arbitrary}\\
&(x \lnot{x} \lnot{x}) \rightarrow \emptyset \tag{$x$ arbitrary}\\
&(\lnot{x} \lnot{x} \lnot{x}) \rightarrow (\lnot x ab) \land (\lnot x \lnot a b) \land (\lnot x \lnot b c) \land (\lnot x \lnot b \lnot c) \tag{$x=F$}\\
&(xxy) \rightarrow (xy \lnot a) \land (abc) \land (a \lnot b c) \land (a \lnot c d) \land (a \lnot c \lnot d) \tag{$(x=T)$ or $(y=T)$}\\
&(\lnot{x}\lnot{x}y) \rightarrow (\lnot x y \lnot a) \land (abc) \land (a \lnot b c) \land (a \lnot c d) \land (a \lnot c \lnot d) \tag{$(x=F)$ or $(y=T)$}\\
&(xx\lnot{y}) \rightarrow (x \lnot y \lnot a) \land (abc) \land (a \lnot b c) \land (a \lnot c d) \land (a \lnot c \lnot d) \tag{$(x=T)$ or $(y=F)$}\\
&(\lnot{x}\lnot{x}\lnot{y}) \rightarrow (\lnot x \lnot y \lnot a) \land (abc) \land (a \lnot b c) \land (a \lnot c d) \land (a \lnot c \lnot d) \tag{$(x=F)$ or $(y=F)$}\\
&(x\lnot{x}y) \rightarrow \emptyset \tag{$x$, $y$ arbitrary}\\
&(x\lnot{x}\lnot{y}) \rightarrow \emptyset \tag{$x$, $y$ arbitrary}
\end{align*}
If $x$ appears in a single clause, we add two new clauses:
\begin{align*}
&\emptyset \rightarrow (xab) \land (xab) \tag{$x$ arbitrary}
\end{align*}

Now we show that the transformations maintain the nesting of the clauses, i.e., they do not introduce crossings. Removing a clause clearly does not affect the embedding. The remaining transformations are of the following type:

\begin{enumerate}[(i)]
\item for given $x$, add $(xab) \land (x \lnot a b) \land (x \lnot b c) \land (x \lnot b \lnot c)$
\item for given $x$ and $y$ appearing in the same clause, add $(xy \lnot a) \land (abc) \land (a \lnot b c) \land (a \lnot c d) \land (a \lnot c \lnot d)$
\item for given $x$, add $(xab) \land (xab)$.
\end{enumerate}

Figure~\ref{figxx} shows how we can make these changes while maintaining proper nesting. In (i) vertices $a$, $b$ and $c$ are placed near $x$, so that $x$ can still be linked to any other clause. In (ii) vertex $a$ is placed near $x$, such that the clause $(xy \lnot a)$ can take the place of the previous clause in which $x$ and $y$ appeared. We place $b$, $c$ and $d$ near $a$ as we did in (i). In (iii) we place $a$ and $b$ near $x$. In this way, $x$ can still be connected to any clause. ~\hfill\qed

\begin{figure}[h!]
  \centering
\includegraphics[scale=0.22]{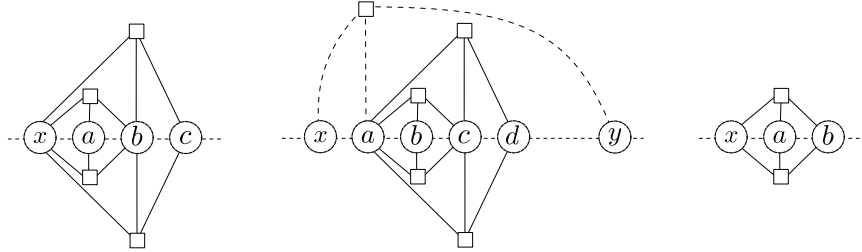}
  \caption{Transformations corresponding to cases (i), (ii) and (iii).} \label{figxx}
\end{figure}
\end{proof}

\vspace{0.2in}

\begin{figure}[h!]
  \centering
\includegraphics[scale=0.35]{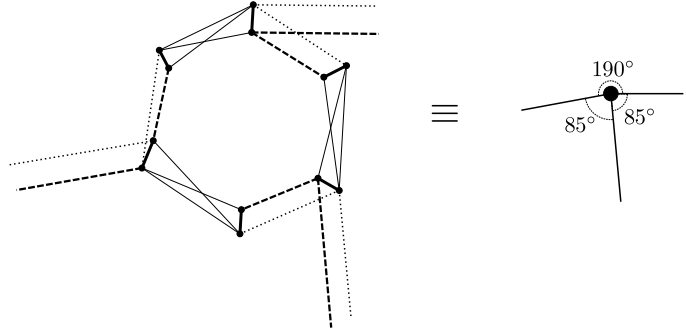}
  \caption{A stretched clause.} \label{fig7}
\end{figure}

\begin{proof} (Lemma~\ref{lema1_})

Both the variable and clause gadgets can be rotated, they can be made arbitrarily small and the bridges connecting them can be arbitrarily narrow. It remains to be shown that edges can emanate from a vertex at any required angle. In a planar circuit, we can move the clauses arbitrarily close to the line on which the variables lie, thereby making the three angles between the bridges of a clause $90^\circ - \varepsilon_1$, $90^\circ - \varepsilon_2$ and $180^\circ + \varepsilon_3$, with $\varepsilon_1$, $\varepsilon_2$, $\varepsilon_3$ positive and arbitrarily close to zero, such that $\varepsilon_1 + \varepsilon_2 - \varepsilon_3 = 0$.

We then show by construction that a clause gadget can be stretched such as to have two angles of $85^\circ$ and one angle $190^\circ$ (Fig.~\ref{fig7}). Since the gadget having these angles can be stretched continuously at any bridge to the symmetric gadget with all angles having $120^\circ$, a clause with angles $90^\circ - \varepsilon_1$, $90^\circ - \varepsilon_2$, $180^\circ + \varepsilon_3$ can clearly be represented. 

For variable gadgets we can prove something stronger: we can represent any angle in the interval $(0^\circ,360^\circ)$ between neighboring bridges. To see this, consider a wire-piece between two neighboring bridges that does not bend at all, in which case the bridges are parallel. To produce a full circle, 16 vertices are sufficient, as seen in Fig.~\ref{fig4}(b). Therefore, if we enforce that the number of vertices between two bridges is greater than 16, we can represent any angle. ~\hfill\qed
\end{proof}

\newpage
\vspace{0.2in}

\paragraph{Irrelevant edges.} 


Let us count the number of vertices in the resulting {\sc madt} instance. Clause gadgets have 12 vertices each, as seen in Fig.~\ref{fig4}(c), therefore the total number of vertices due to clauses is $12 n_c$. The number of vertices in variable gadgets is $3 n_c(N+2)$ (for all $3 n_c$ bridges we have 2 base vertices and $N$ vertices separating it from the next bridge in clockwise order). The number of all vertices in the construction is thus $18 n_c + 3 n_c N$ which is strictly smaller than $4 n_c N$, assuming that $N>18$.

Since the whole construction is connected and non-irrelevant edge weigths are at most 1, the longest possible distance is smaller than $4 n_c N$. Thus, we can set the weights of irrelevant edges $\sigma = (4 n_c N)^3 > \mathcal{W}(T)$. In this way the irrelevant edges never contribute to the cost, therefore we can simply ignore them. Note that the weights $\sigma$ violate the triangle inequality.\\

\vspace{0.2in}
\begin{proof} (Lemma~\ref{thm5})

Suppose a variable is in impure state. Start with a dashed edge and follow the loop in clockwise order. At some point we switch to dotted edges. Where this happens, we have a local structure that we call \textit{bubble}. There can be two different types of bubble, depending on whether the transition occurs at a bridge or somewhere else in the wire-piece. As we continue on the loop, we have to switch back to dashed edges somewhere. Where this happens, locally we have a \textit{hole}. Figure~\ref{fig9} shows these local features. It is easy to see that if a variable gadget is impure, it has to have at least one hole and at least one bubble. What we show is that we can remove a hole and a bubble while decreasing the cost of the triangulation. When all holes and bubbles are removed, the triangulation is pure.

\begin{figure}[h!]
  \begin{center}
    \subfigure[]{\label{fig9a}\includegraphics[trim = 0mm -60mm 0mm 0mm, clip,scale=0.2]{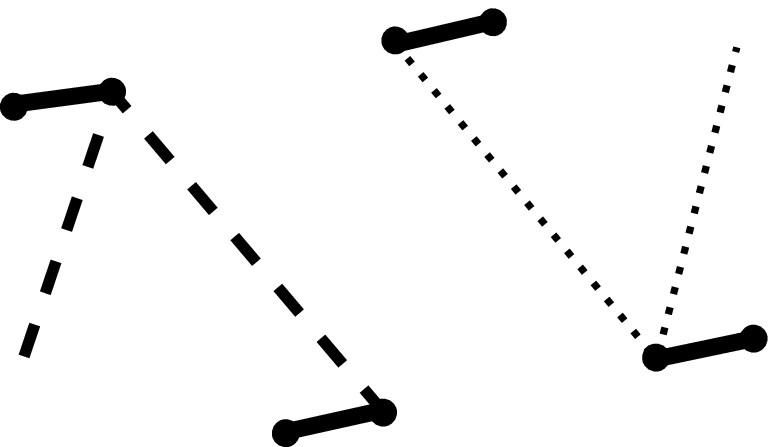}}
    \subfigure[]{\label{fig9b}\includegraphics[trim = -30mm -70mm 0mm 0mm, scale=0.2]{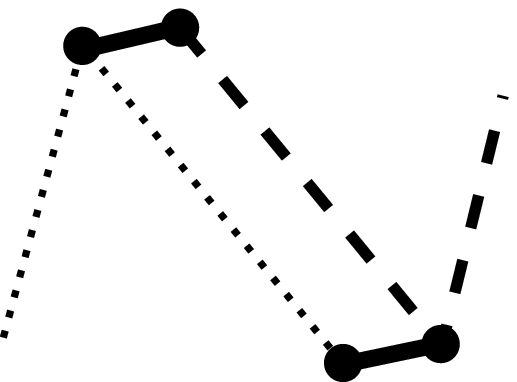}}
    \subfigure[]{\label{fig9c}\includegraphics[trim = -40mm -20mm 0mm 0mm, scale=0.2]{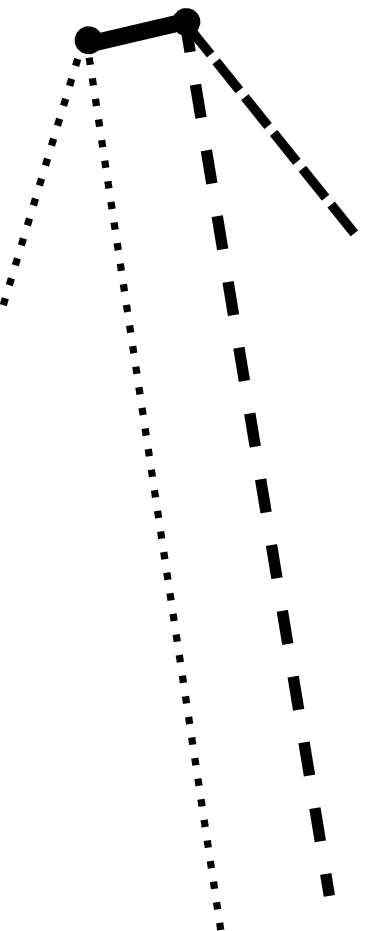}}
    \subfigure[]{\label{fig9d}\includegraphics[trim = -50mm 0mm 0mm 0mm, scale=0.2]{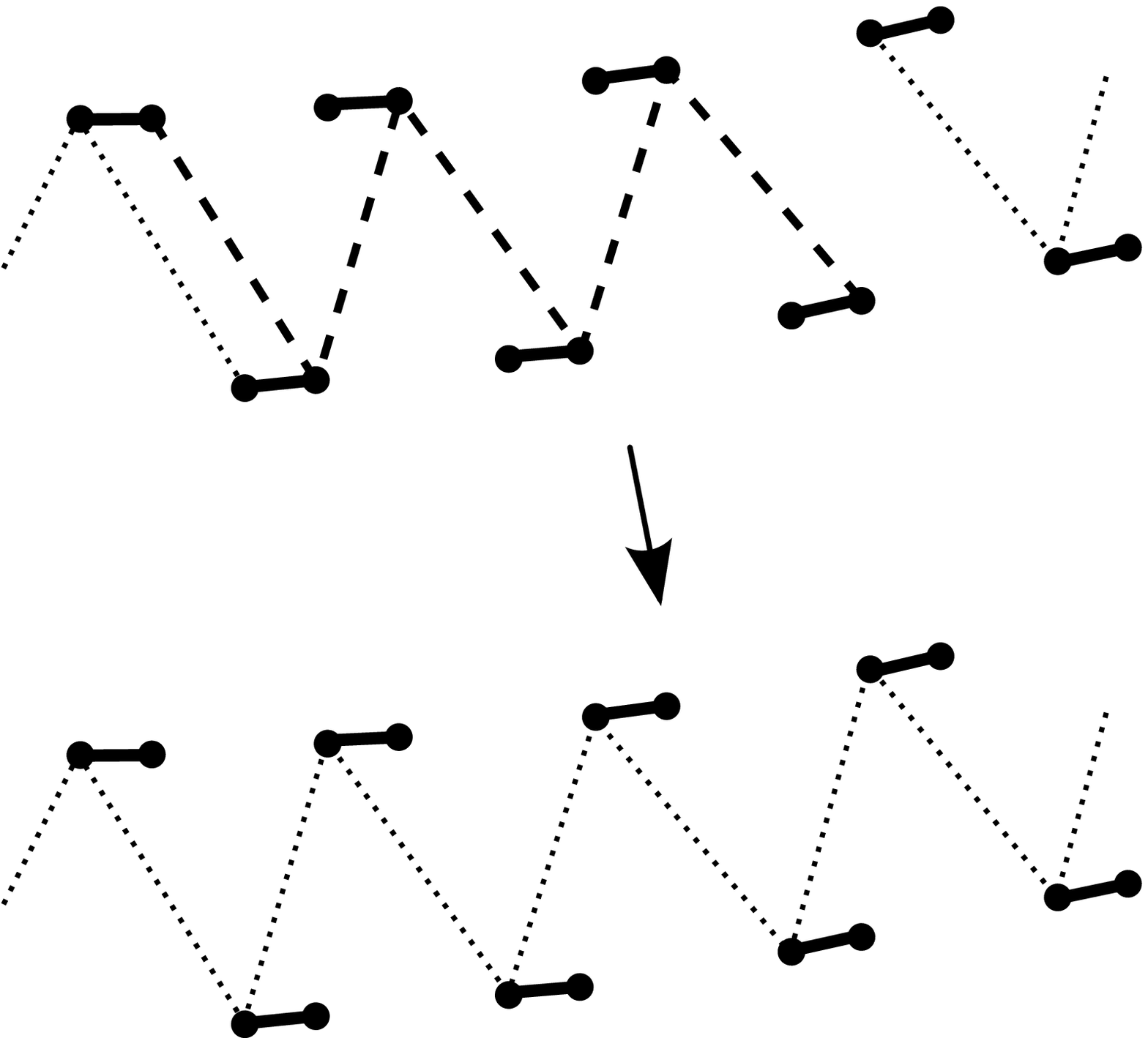}}
  \end{center}
  \caption{(a) Hole. \quad (b) Bubble. \quad (c) Bubble at bridge. \quad (d) Removing a bubble and a hole.}
  \label{fig9}
\end{figure}



We consider a sequence consisting of a hole, a pure piece of wire and a bubble (of either type). The pure piece of wire might have several attached bridges along the way. We flip each flippable edge of the pure wire-piece (including any bridges along the way), thereby absorbing the hole and the bubble at the ends (Fig.~\ref{fig9d}). Now we look at the change in cost due to this operation. We count first the distances that might have become larger, then we count the distances which provably became smaller.

Take any two vertices and the shortest path between them. If a pure wire-piece contained in this path is flipped, the distance between the two endpoints can increase by at most $2$ (we can simulate the old path with the new path and two extra edges of length $1$ at both ends). If a bridge along the path is flipped (or partially deleted, if a bridge-bubble was removed), this can also increase any distance by at most $2$ (we can simulate the old bridge using the new bridge and two extra edges at the bridge endpoints of length $1$ each). If a bridge is flipped, this can force a change within the clause to which it is connected. In Fig.~\ref{fig4}(c) we can verify that in any triangulation of a clause the distance between two vertices is less than 5. Therefore, a single clause can increase a path that somehow intersects it by less than 5 during this transformation. The total number of clauses is $n_c$ and the number of bridges $3 n_c$, therefore the penalty on any distance between two points is at most $2 + 3 n_c \cdot 2 + n_c \cdot 5 = 11 n_c + 2$. The number of distances is less than ${4 n_c N \choose 2}$, therefore the total penalty due to removing one bubble and one hole is smaller than $88 {n_c}^3 N^2 + 16 {n_c}^2 N^2 < 100 {n_c}^3 N^2$ (assuming $n_c > 1$). 

Now we look at distances that provably decrease with the transformation. We can assume that between two neighboring bridges the variable gadget contains a single hole. If there were more, some vertices would be isolated and we would have to cross an irrelevant edge, incurring a cost of $\sigma$. In this case, removing the hole would obviously decrease the cost by making the construction connected using only non-irrelevant edges. 

Consider now such a hole on the wire between two bridges (Fig.~\ref{fig11}). Denote by $n_1$ the number of vertices between the previous bridge (in clockwise order) and the hole and by $n_2$ the number of vertices between the hole and the next bridge. Assume w.l.o.g. that $n_1 \leq n_2$. Depending on the state of the bridges, $n_1+n_2$ can take the values $N$, $N+1$ or $N+2$. It follows that $n_2 \geq \frac{N}{2}$ and $n_1 \leq \frac{N}{2}+1$.  Among the $n_2$ vertices after the hole, consider the $\frac{N}{8}$ vertices closest to the hole and denote this set by $A$. On the other side of the other bridge denote the $\frac{N}{8}$ vertices closest to the bridge by $B$ (Fig.~\ref{fig11}(a)). Assume, for now, that $B$ is free of holes. We will treat the case when $B$ has a hole, later. To make things simpler, we choose $N$ to be divisible by 16. The following small lemma will help us with the computations. Its validity can be seen by a simple analysis of cases (see Fig.~\ref{fig5}(a) and Fig.~\ref{fig9d}).

\begin{lemma} 
\label{lemt}
If $M$ is the longest distance between two vertices in a hole-free wire-piece with $k$ vertices (possibly containing a bubble), then $\frac{k}{2} \leq  M \leq \frac{k}{2} + 2$ holds. ~\hfill\qed
\end{lemma}

\begin{figure}[h!]
  \centering
    \subfigure[]{\label{fig:fig16a}\includegraphics[scale=0.37]{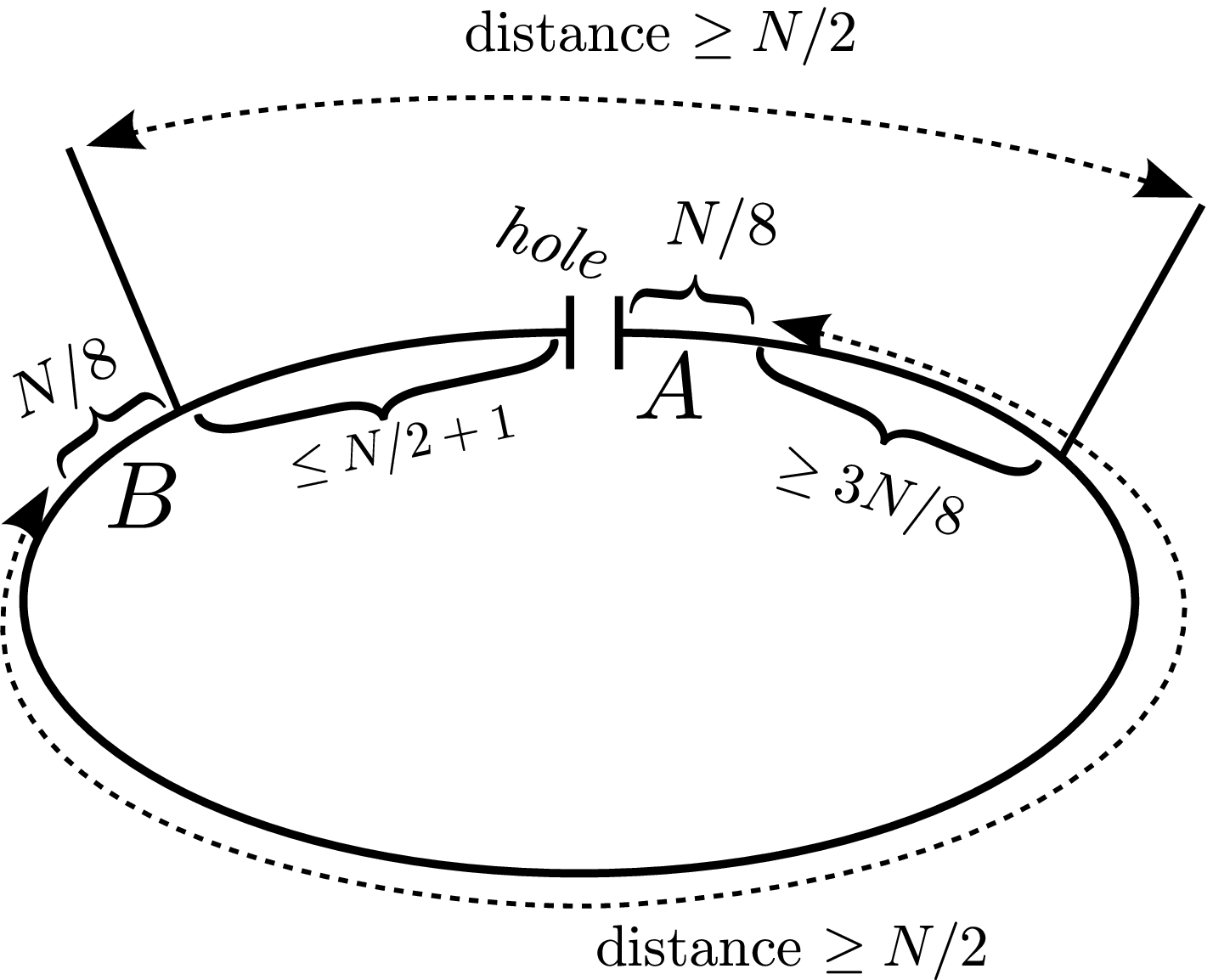}}
    \subfigure[]{\label{fig:fig16b}\includegraphics[trim = -20mm 0mm 0mm 0mm, clip, scale=0.37]{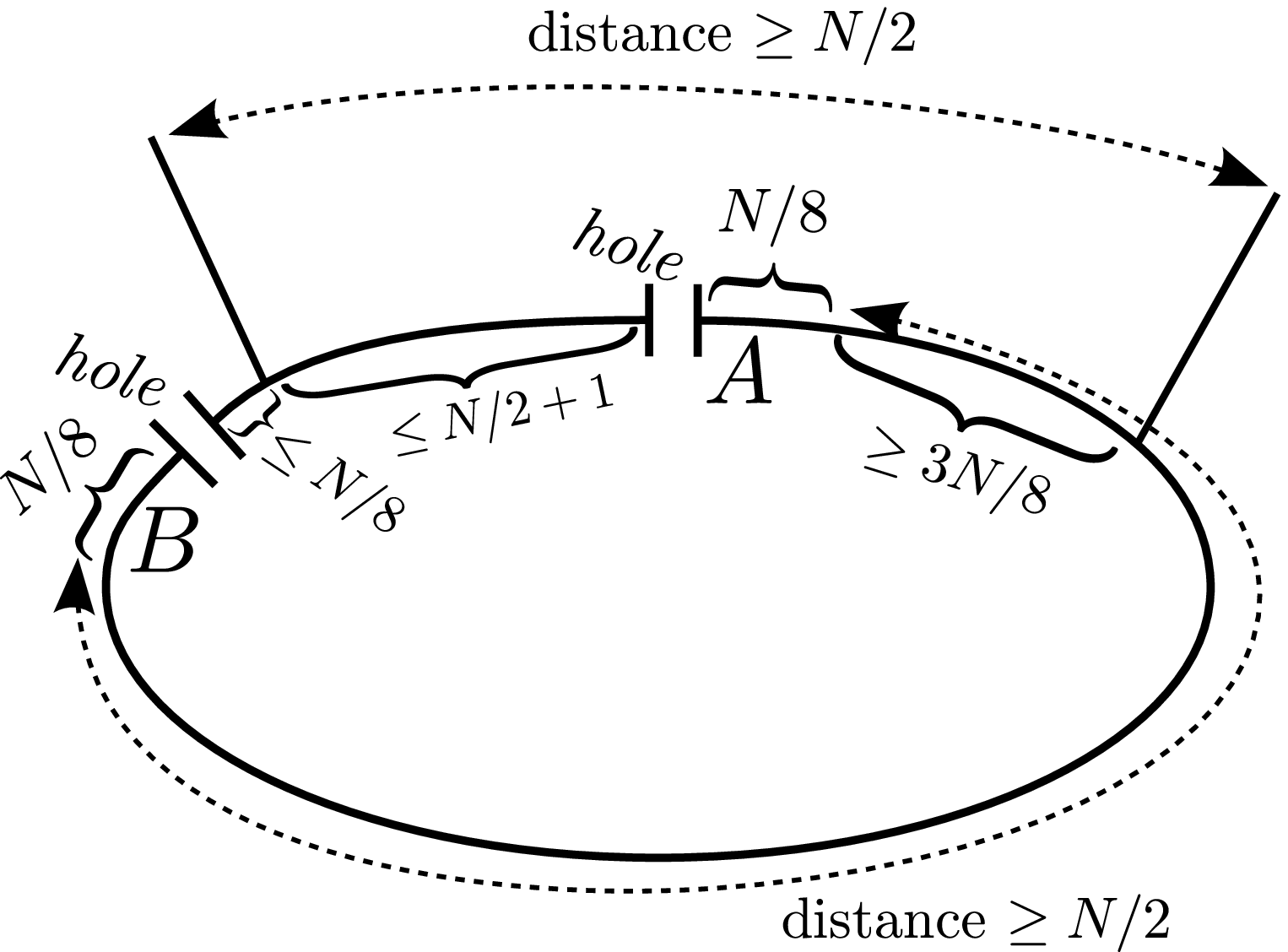}}
  \caption{Sketch of a variable gadget. $A$ and $B$ are brought closer by the removal of the hole(s).} \label{fig11}
\end{figure}

We know that all variables have at least two bridges (due to R2) and all bridges of a variable are connected to different clauses (due to R1). Let us look at the minimum distance from a vertex in $A$ to a vertex in $B$ while the hole is still there. In any direction we have to go through a wire containing at least $N$ vertices, therefore the distance is at least $\frac{N}{2}$ (due to Lemma~\ref{lemt}). When the hole is removed, the distance between a vertex in $A$ and a vertex in $B$ is at most $\frac{3N}{8} + 3$.

If there was a hole within the $\frac{N}{8}$ vertices that we labeled as $B$, we take instead the $\frac{N}{8}$ vertices immediately before this hole and label them as $B$ instead (Fig.~\ref{fig11}(b)). In this case we can still claim that the distances between A and B are at least $\frac{N}{2}$ before the transformation. Now we remove both holes (and the two corresponding bubbles), potentially inflicting twice the penalty which we bounded from above by $100 {n_c}^3 N^2$. After removing both holes, distances between $A$ and $B$ are at most $\frac{7N}{16} + 3$. We get a decrease in cost of at least $\frac{N}{16} - 3$ for $\frac{N}{8} \cdot \frac{N}{8}$ pairs.

If we enforce $N > 5 \cdot 10^5 {n_c}^3$, we get a net decrease in cost due to the removal of the hole(s). Therefore, we can transform any impure triangulation into a pure one of lower cost. ~\hfill\qed
\end{proof}

\newpage
\begin{proof} (Lemma~\ref{thm6})

The high level idea is the following: we first count the distances that can be smaller in $T_\textrm{nonSAT}$ than in $T_\textrm{SAT}$ and we bound their contribution to $\mathcal{W}$. Then we count those distances that are provably smaller in $T_\textrm{SAT}$ than in $T_\textrm{nonSAT}$ and we add up the differences. The crucial fact that makes the proof possible is the following: $T_\textrm{nonSAT}$ has at least one clause with all three literals \emph{false}. For this clause, crossing the gadget from one bridge to another has a cost of at least $1+2\varepsilon$. In $T_\textrm{SAT}$ clause crossings have cost at most $4 \varepsilon$ (Fig.~\ref{fig12}). Using the fact that each clause crossing participates in $\Omega(N^2)$ distances, given bounds on $N$ and $\varepsilon$ we obtain the required bound on the difference between the costs.

Figure~\ref{fig12} shows the optimum triangulation of the clause gadgets in each possible assignment, ignoring symmetric cases. We are interested in the distances between the endpoints of the three bridges in the clause. These are summarized in the bottom row of Fig.~\ref{fig12}. The triangulations are optimal in the sense that no other triangulation can achieve a lower distance between any of the bridge endpoints. These will be the triangulations used in $T_\textrm{SAT}$, but in $T_\textrm{nonSAT}$ we will implicitly consider other triangulations of the clauses as well. Intuitively, it is clear that nonsatisfied clauses $\{F, F, F\}$ are costlier to cross than satisfied ones, and indeed, this is what makes our reduction possible. Now we make this intuition more precise. 

\begin{figure}[h!]
  \centering
\includegraphics[scale=0.24]{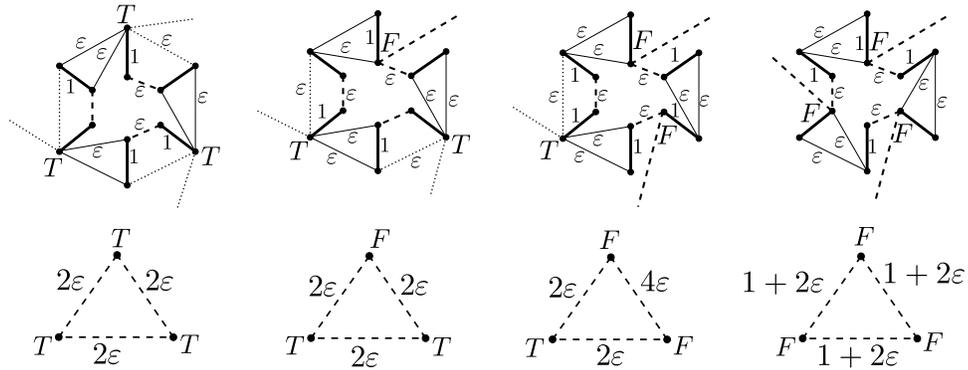}
  \caption{(top) Optimal triangulation of a clause. \quad (bottom) Cost of crossing a clause.} \label{fig12}
\end{figure}

Every distance between two vertices is of one of the following types: (i) between vertices of the same clause, (ii) between vertices of the same variable, (iii) between a vertex from a clause and a vertex from a variable, (iv) between vertices of different clauses, and (v) between vertices of different variables.

We go through the five types of distances and denote their contribution to the cost by $\mathcal{W}_1$, \dots, $\mathcal{W}_5$. In each of the five cases we want to compare the cost of $T_\textrm{SAT}$ with the cost of $T_\textrm{nonSAT}$. 

(i) Within a single clause, even in the most unfavorable triangulation, the distance between any two vertices is less than $5$, therefore $\mathcal{W}_1 < {n_c}{12 \choose 2}5 = 330 n_c$.

(ii) Variable gadgets in the two different states are isomorphic, therefore $\mathcal{W}_2(T_\textrm{SAT}) = \mathcal{W}_2(T_\textrm{nonSAT})$.

(iii) There are less than $(4 n_c N)(12 n_c)$ such distances. If we look at one shortest path as we move from a satisfying to a non-satisfying assignment, the path length can decrease by at most $1$ at both endpoints. Variable-crossings and bridge-crossings along the way maintain their length and clause-crossings can decrease by at most $2\varepsilon$ each (compare crossing costs of clauses in Fig.~\ref{fig12}). Therefore $\mathcal{W}_3(T_\textrm{SAT}) - {\mathcal{W}_3}(T_\textrm{nonSAT}) < (2+6 n_c\varepsilon)(4 n_c N)(12 n_c)$, which, assuming $\varepsilon < \frac{1}{6 n_c}$, is less than $144 {n_c}^2 N$.

(iv) By a similar argument, $\mathcal{W}_4(T_\textrm{SAT}) - {\mathcal{W}_4}(T_\textrm{nonSAT}) \linebreak < (2+6 n_c\varepsilon)(12 n_c)(12 n_c) < 432 {n_c}^2$, assuming $\varepsilon < \frac{1}{6 n_c}$.

(v) This part is the crucial one, since it contributes the highest order term in $N$ to the cost. Our goal is to show that $\mathcal{W}_5(T_\textrm{nonSAT}) - {\mathcal{W}_5}(T_\textrm{SAT}) = \Theta(N^2)$, outweighing the other four differences which are all $\O(N)$.

Let us look at the distance between two vertices, $p_x$ and $p_y$ from different variable gadgets (Fig.~\ref{fig14}). Let $d(p_x,p_y)$ be their distance in $T_\textrm{SAT}$ and $d'(p_x,p_y)$ their distance in $T_\textrm{nonSAT}$. 

\begin{figure}[h!]
  \centering
\includegraphics[scale=0.35]{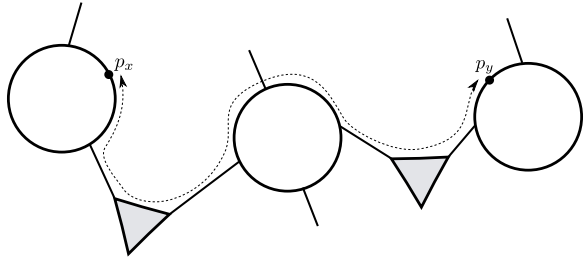}
  \caption{Shortest path between $p_x$ and $p_y$. Variables appear as circles, clauses as triangles.} \label{fig14}
\end{figure}

We denote variable gadgets as $V_1, \dots, V_{n_v}$ and we write the cost due to distances between vertices from different variables as:
\[
\mathcal{W}_5(T_\textrm{SAT}) = \displaystyle\sum_{\substack{p_x,p_y:\\p_x \in V_i, \; p_y \in V_j\\1 \leq i < j \leq n_v}}{d(p_x,p_y)}.
\]

In Fig.~\ref{fig15} we see that every vertex has a natural \textit{neighbor}, the vertex to which it is connected by a thick solid edge. We denote the neighbors of $p_x$ and $p_y$ as $\overline{p_x}$ and $\overline{p_y}$, respectively. We define the distance between two pairs of neighboring points as follows:
\[
d\bigl([p_x \overline{p_x}], [p_y \overline{p_y}]\bigr) = d(p_x, p_y) + d(\overline{p_x}, p_y) + d(p_x, \overline{p_y}) + d(\overline{p_x}, \overline{p_y}).
\]

\begin{figure}[h!]
  \centering
\includegraphics[scale=0.2]{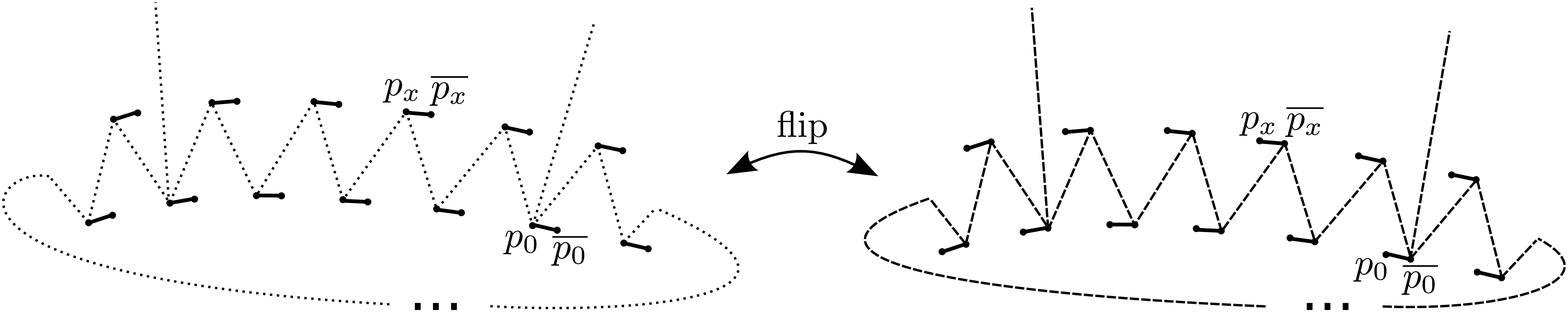}
  \caption{Vertices in a bridge-to-bridge portion of a variable gadget.} \label{fig15}
\end{figure}

One vertex out every pair of neighbors is a \textit{leaf vertex}, in the sense that a path from that vertex to any other vertex goes through its neighbor. In Fig.~\ref{fig15}, left, $\overline{p_x}$ and $\overline{p_0}$ are leaf vertices, but when the variable is flipped into the other pure state (Fig.~\ref{fig15}, right), the situation reverses and $p_x$ and $p_0$ become leaves. We can simplify the distance between pairs of points as follows:

\[
d\bigl([p_x \overline{p_x}], [p_y \overline{p_y}]\bigr) = 4\;\upphi(p_x, p_y) + 4,
\]

where $\upphi$ is the distance between the non-leaf members of both pairs, or more precisely:

\[
\upphi(p_x, p_y) = \min \bigl( d(p_x, p_y), \; d(\overline{p_x}, p_y), \; d(p_x, \overline{p_y}), \; d(\overline{p_x}, \overline{p_y}) \bigr).
\]

Now we can write the relevant part of the cost in terms of $\upphi$:
\[
\mathcal{W}_5(T) = \displaystyle\sum_{\substack{p_x,p_y:\\p_x \in v_i, \; p_y \in v_j\\1 \leq i < j \leq n_v}}{\bigl(\upphi(p_x,p_y) + 1\bigr)}.
\]

Let $\upphi(p_x,p_y)$ denote the distance defined above in $T_\textrm{SAT}$ and $\upphi'(p_x,p_y)$ the corresponding distance in $T_\textrm{nonSAT}$. We want to bound $\upphi - \upphi'$ from above.
Let us decompose $\upphi(p_x,p_y)$ into components. Remember that $\upphi$ is the distance between two non-leaf points, i.e., the length of the shortest path between them. Such a path goes from $p_x$ to a bridge, then crosses a number of bridges, clauses and variables, arrives to the target variable, and goes from the bridge to $p_y$. Observe that the first and last components (endpoint to bridge) do not change with the flipping of a variable. This can be seen in Fig.~\ref{fig15}: the distance $d(p_x, p_0)$ on the left and the distance $d(\overline{p_x}, \overline{p_0})$ on the right are equal. Variable-crossing costs do not change either, a variable bridge-to-bridge portion always has distance $\frac{N}{2} + 1$ and neither do bridge-crossings which always cost $\varepsilon$. 

The only difference in cost between $\upphi$ and $\upphi'$ is due to clause crossings. Whereas $T_\textrm{SAT}$ contains only clauses of the type $\{T,T,T\}$, $\{T,T,F\}$, $\{T,F,F\}$, in $T_\textrm{nonSAT}$ we have at least one $\{F,F,F\}$ clause. Thus, according to Fig.~\ref{fig12}, the maximum cost of a crossing in $T_\textrm{SAT}$ is $4\varepsilon$ and the minimum cost of a crossing in $T_\textrm{nonSAT}$ is $2\varepsilon$. A shortest path can cross each clause only once, otherwise there would exist a shortcut. Since there are $n_c$ clauses in total, we obtain the bound:

\[
\upphi(p_x,p_y) \leq \upphi'(p_x,p_y) + 2 n_c \varepsilon.
\]

In $\mathcal{W}_5$ we have at most ${3 n_c \choose 2}(N + 2)^2$ distances, each of which can be shorter by at most $2 n_c \varepsilon$ in $T_\textrm{nonSAT}$ than in $T_\textrm{SAT}$ (provided that we group distances four-by-four as explained above and we average over the groups). The number of distances (assuming $N \geq 12 {n_c}^2$) is less than $4 {n_c}^2N^2$.

Now let us look at distances that are provably larger in $T_\textrm{nonSAT}$ than in $T_\textrm{SAT}$. We know that there is at least one clause crossing that has cost $1+2\epsilon$ in $T_\textrm{nonSAT}$ and cost at most $4\varepsilon$ in $T_\textrm{SAT}$. This clause crossing is thus at least $1-2\varepsilon$ costlier in $T_\textrm{nonSAT}$ than in $T_\textrm{SAT}$. For every clause crossing there are at least $\frac{N}{4} \cdot \frac{N}{4}$ shortest paths going through that crossing, regardless of the states of the variables, i.e., both in $T_\textrm{nonSAT}$ and in $T_\textrm{SAT}$. This fact is illustrated in Fig.~\ref{fig16} (sets $A$ and $B$). In this way we get that there are at least $(\frac{N}{4})^2$ distances that contribute at least $1-2\varepsilon$ more to $\mathcal{W}_5 (T_\textrm{nonSAT})$ than to $\mathcal{W}_5 (T_\textrm{SAT})$.

\begin{figure}[h!]
  \centering
\includegraphics[scale=0.70]{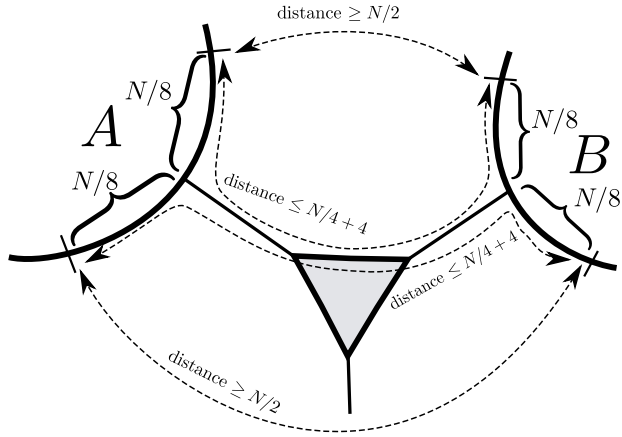}
  \caption{Shortest paths that have to go across a given clause crossing.} \label{fig16}
\end{figure}

Now we have all the ingredients to compare $\mathcal{W} (T_\textrm{nonSAT})$ and $\mathcal{W} (T_\textrm{SAT})$:

\begin{align*}
\mathcal{W} (T_\textrm{nonSAT}) - \mathcal{W} (T_\textrm{SAT}) = & \;\; \mathcal{W}_1 (T_\textrm{nonSAT}) - \mathcal{W}_1 (T_\textrm{SAT}) + \mathcal{W}_2 (T_\textrm{nonSAT}) - \mathcal{W}_2 (T_\textrm{SAT})\\
& \; + \mathcal{W}_3 (T_\textrm{nonSAT}) - \mathcal{W}_3 (T_\textrm{SAT}) + \mathcal{W}_4 (T_\textrm{nonSAT}) - \mathcal{W}_4 (T_\textrm{SAT})\\
& \; + \mathcal{W}_5 (T_\textrm{nonSAT}) - \mathcal{W}_5 (T_\textrm{SAT}) \\
\geq & \; - 330 n_c - 144 {n_c}^2 N- 432 {n_c}^2 - 8{n_c}^3 N^2 \varepsilon + (\frac{N}{4})^2 (1-2\varepsilon)\\
\geq & \; \frac{N^2}{32}.\\
 \tag{assuming $N>5 \cdot 10^5 {n_c}^3$ and $\varepsilon<\frac{1}{N^2}$}
\end{align*}
~\hfill\qed
\end {proof}

\newpage
\begin{proof} (Lemma~\ref{thmxy})

We go through the same computations as for Lemma~\ref{thm6} and use the fact that between two satisfying assignments a clause crossing can change cost by at most $2\varepsilon$.~\hfill\qed
\end{proof}

\vspace{0.1in}

\paragraph{Baseline Triangulation.} 

Our goal is to generate $\mathcal{W}^\star$ somewhere in the gap between the costs of satisfying and non-satisfying triangulations. It would be sufficient to generate a triangulation corresponding to a satisfying assignment and add $150 {n_c}^2 N$ to its cost. We do not even know, however, whether a satisfying assignment exists.

Instead, we construct a simpler triangulation that we call \textit{baseline}: we assign to each variable an arbitrary truth value and triangulate the variable gadgets and attached bridges accordingly. Then we replace each clause with the baseline gadget of Fig.~\ref{figbase}, connecting the three vertices of the triangle to the bridge endpoints that that were supposed to connect to that clause. We note that many other configurations would work similarly well as a baseline gadget. For the described construction we can compute $\mathcal{W}(T_\textrm{baseline})$ using an all-pairs shortest path algorithm. \\

\vspace{0.05in}

\begin{proof} (Lemma~\ref{thmxyz})

The computations are similar to those in the previous proofs (we look at the five different types of distances):
\begin{enumerate}[(i)]
\item Within a single clause of $T_\textrm{baseline}$ the distance between any two vertices is less than $2$, therefore $\mathcal{W}_1(T_\textrm{baseline}) < n_c{12 \choose 2}2 = 132 n_c$.

\item Here also $\mathcal{W}_2(T_\textrm{baseline}) = \mathcal{W}_2(T_\textrm{SAT})$.

\item Here also $\bigl| \mathcal{W}_3(T_\textrm{baseline}) - {\mathcal{W}_3}(T_\textrm{SAT})\bigr| < (2+6c\varepsilon)(4 n_c N)(12 n_c)<144 {n_c}^2 N$.

\item Here also $\bigl| \mathcal{W}_4(T_\textrm{baseline}) - {\mathcal{W}_4}(T_\textrm{SAT})\bigr| < (12 n_c)(12 n_c)(2+6 n_c \varepsilon) < 432 {n_c}^2$.

\item Here also, clause crossings can change by at most 2$\varepsilon$, therefore $\bigl| \mathcal{W}_5(T_\textrm{baseline}) - {\mathcal{W}_5}(T_\textrm{SAT})\bigr| < 8{n_c}^3 N^2 \varepsilon$.
\end{enumerate} 

Overall we find that $\bigl|\mathcal{W}(T_\textrm{baseline})-\mathcal{W}(T_\textrm{SAT})\bigr| \leq 150 {n_c}^2 N$. ~\hfill\qed

\begin{figure}[h!]
  \centering
\includegraphics[scale=1.40]{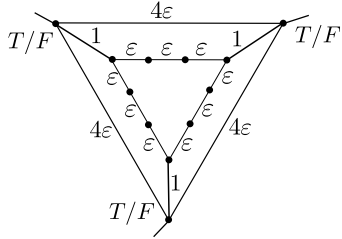}
  \caption{Clause gadget in baseline triangulation.} \label{figbase}
\end{figure}

\end{proof}

\end{document}